\documentclass[twocolumn,
prd,secnumarabic,nobalancelastpage,amsmath,amssymb,superscriptaddress,
nofootinbib]{revtex4-2}

\usepackage{graphicx} 
\usepackage{xcolor}
\usepackage{multirow}
\newcommand{\mE}{{\mathcal{E}}}
\newcommand{\mB}{{\mathcal{B}}}

\newcommand{\req}[1]{Eq.\,(\ref{#1})} 
\newcommand{\rs}[1]{section~\ref{#1}} 
\newcommand{\rf}[1]{figure\,\ref{#1}} 
\newcommand{\fracg}{\frac{g}{2}}

\definecolor{blue}{rgb}{0,0.2,0.8}

\begin{document}

\title{Singular properties of QED vacuum response\\ 
to applied quasi-constant electromagnetic fields}

 \author{Stefan Evans}
 \email{evanss@arizona.edu}
 \author{Johann Rafelski}
 \affiliation{Department of Physics,
 The University of Arizona,
 Tucson, Arizona, 85721 USA}

\begin{abstract}
Employing the Bogoliubov coefficient summation method and introducing the gyromagnetic ratio $g\neq 2$ we derive an explicit functional form of $\mathfrak{Im}V^\mathrm{EHS}_g$, the imaginary part of Euler-Heisenberg-Schwinger (EHS) type effective action. We show that $\mathfrak{Im}V^\mathrm{EHS}_g$ is periodic in $g$ for any (quasi-)constant electromagnetic field configuration, and equal to the imaginary part obtained using a periodic in $g$ Ramanujan integrand in the proper time representation of $V^\mathrm{EHS}_g$. This validates the Ramanujan representation of $V^\mathrm{EHS}_g$ for both real and imaginary parts and allows writing the effective action in a suitably modified Schwinger proper time format. As a function of the ratio $b/a$ between $\mB \to b$ and $\mE\to a$ covariant generalizations of EM fields, we explore the singular properties of $\mathfrak{Im}V^\mathrm{EHS}_g$ at $g=2\pm 4k, k=0,\pm1,\pm2\ldots$ involving the pseudoscalar $ab\equiv \vec\mE\cdot\vec\mB$ in perturbative and nonperturbative behavior. We study the $e^-e^+$-decay vacuum instability, incorporating the physical value of $g-2$ vertex diagrams when summing infinite irreducible loops. We obtain an effective expansion parameter $\chi_b=\alpha b/2a$ ($\alpha=e^2/4\pi$), characterizing the onset of nonperturbative in $g-2$ suppression of vacuum instability. We demonstrate the $\chi_b$ domains for which perturbative expansion in $\alpha$ breaks down: The EM vacuum subject to critical electric field strength is stabilized in magnetic-dominated \lq magnetar\rq\ environments. Considering separately the case of $\mE$ and $\mB$ fields, we generalize to all $g$ the temperature representation of the $V^\mathrm{EHS}_g$ effective action. 
\end{abstract}

\maketitle

\section{Introduction}

The response by virtual electron-positron $e^-e^+$-pairs to the action of an externally applied nearly constant {\it i.e.\/} quasi-constant electro-magnetic (EM) field $\mE, \mB$ has been explored in the seminal work by Euler-Heisenberg-Schwinger (EHS)~\cite{Heisenberg:1935qt, Weisskopf:1996bu, Schwinger:1951nm}, the effective QED action $V^\mathrm{EHS}$. The imaginary part $\mathfrak{Im}V^\mathrm{EHS}$  relates to the probability of the field filled vacuum state to decay into $e^-e^+$-pairs.

The EHS effective action is built on solutions to the Dirac equation, with fixed physical quantities mass $m$, charge $e$, and {\em magnetic moment} described in terms of  gyromagnetic ratio $g=2$. Beyond this framework, these physical quantities are modified by higher order QED interactions~\cite{Ritus:1970, Jancovici:1970ep, Newton:1971pq, Constantinescu:1972qe, Tsai:1974id, Narozhnyi:1979at, Morozov:1981pw, Loskutov:1981bk, Gusynin:1998nh, Machet:2015swa, Ferrer:2015wca, DiPiazza:2021szp}, which in turn can be implemented as corrections to the EHS result. An example of such corrections to the EHS result is the perturbative two-loop action of Ritus~\cite{Ritus:1975cf, Dittrich:1985yb, Fliegner:1997ra, Kors:1998ew, Dunne:1999vd}.

Our objective is a nonperturbative implementation of anomalous magnetic moment $g\neq 2$ in $V^\mathrm{EHS}$, creating $V^\mathrm{EHS}_g$. By incorporating $g\neq2$ via solutions to the relativistic quantum wave equations used to derive the effective action, each virtual particle excitation is thus prescribed its anomalous magnetic moment. This produces a resummation of a class of vertex diagrams to infinite irreducible loop order.

There has been extensive effort based on the Schwinger proper time formulation to implement anomalous magnetic moment in $V^\mathrm{EHS}_g$~\cite{PauliTerm, Dittrich:1977ee, Lav85, Kruglov:2001dp}. Formally the proper time method seems to apply to any value of $g$. However, a closer look at the form of the integrand reveals that the usual method for implementing $g\neq2$ in the proper time representation converges only for $|g|\leq 2$~\cite{Labun:2012jf}. This inspires our effort to obtain using a different method a result for all $g$, allowing for point particles such as the electron where the value $|g|> 2$ matters. 

We apply a method, presented before in our study of an inhomogeneous Sauter step~\cite{Evans:2022fsu}, to obtain $V^\mathrm{EHS}_g$ for any value of $g$ in any quasi-constant EM field configuration. Our approach relies on a constructive Bogoliubov coefficient method developed by Nikishov~\cite{Nikishov:1979ez} and recently elaborated by Kim, Lee and Yoon~\cite{Kim:2008yt}. We follow this work and use a second order fermion formulation of the Dirac equation. Kim et al. considered already the $g=2$ and $g=0$ cases, and we extend it to arbitrary $g$, as we have done in the case of the electric Sauter step~\cite{Evans:2022fsu}.

Allowing for $g\neq2$ (or $g\neq0$) implies a nonperturbative summation of certain classes of diagrams,~\rf{fig:sum}, which accompanies a nonperturbative summation in external EM fields. A new feature based on a double nonperturbative evaluation of $V^\mathrm{EHS}_g$ becomes evident: with each successive order in external photon line summation ($V^\mathrm{EHS}$, top~\rf{fig:sum}), another internal vertex is summed to enclose it ($V^\mathrm{EHS}_g$, bottom~\rf{fig:sum}). Thus just like EHS sums all external fields, $V^\mathrm{EHS}_g$ does the same with the vertex correction to the $g$-factor.

Our result recovers the features from prior work: periodicity as a function of $g$ in pure magnetic $\mB$ fields~\cite{Rafelski:2012ui}, reconfirmed for pure electric $\mE$ fields~\cite{Evans:2022fsu}. Both pure $\mE$ and $\mB$ cases exhibit actions that are peaked yet smooth and differentiable at $g=2$. Here we prove conjectured singular properties with cusp structure at $g=2$ considering $V^\mathrm{EHS}_g$ to all orders in $\vec\mE\cdot\vec\mB$ when the two complementary summations are carried out to infinite order. This requires $\mE$ and $\mB$ fields to have common nonvanishing parallel components. While the key parameter $\vec\mE\cdot\vec\mB$ is a pseudoscalar, $V^\mathrm{EHS}_g$ being even in powers of $\vec\mE\cdot\vec\mB$ is conserving  parity symmetry.

\begin{figure}[bt]
\centering 
\includegraphics[width=0.99\columnwidth]{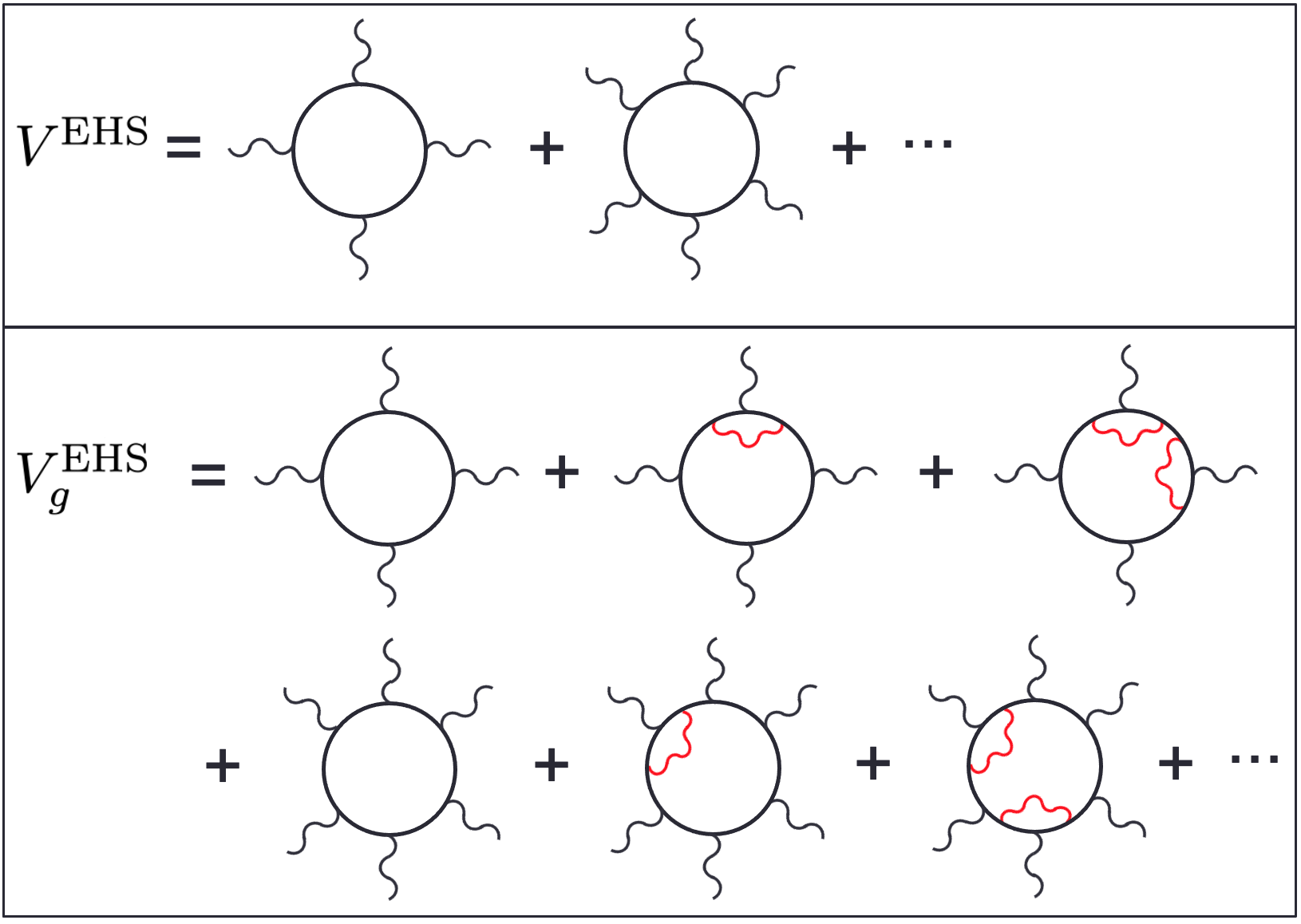}
\caption{$V^\mathrm{EHS}$ and $V^\mathrm{EHS}_g$ diagram summations: Vertex corrections (red) comprise the $g-2$ corrections to all orders. Virtual photon loops enclosing at least two photon lines are not included. 
\label{fig:sum}}
\end{figure}

Beyond the mathematical proof of the cusp, another objective of this work is to understand the EHS particle production dependence on $g$. At $g=2$ the EHS result implies that presence of a strong magnetic field amplifies this effect seen in a pure $\mE$-field. However, in magnetically dominated environments the anomalous magnetic moment adds an extra nonperturbative effect that becomes important when the smallness parameter $\chi_b=\alpha b/2a>1$ (\req{chiB}),  in covariant generalization $\mB \to b$ and $\mE\to a$. It is notable that the reducible QED loop expansion is governed by series in $\alpha^n$, while the magnetic moment expansion is a series in $\chi_b^n$. As a consequence,  the opposite is to be expected allowing for the physical value of electron magnetic moment: suppression of EHS particle production contrary to the expected enhancement.

Our presentation is organized as follows:
In~\rs{KGPfirst} we briefly summarize prior $V^\mathrm{EHS}_g$ work valid for $|g|\leq 2$~\cite{Kruglov:2001dp}. This approach is based on the Schwinger proper time formulation. In~\rs{pseduoNonpert} we derive $V^\mathrm{EHS}_g$ for any $g$. In order to obtain an ab-initio result valid in the domain $|g|>2$ we proceed as follows:\\
a) In~\rs{KGBpseudo} we solve the second order Klein-Gordon-Pauli equation with a spin $g$-factor $g\neq2$. This generalizes the $g=\pm 2$ solution to the Dirac equation used by Heisenberg and Euler~\cite{Heisenberg:1935qt} and allows for $\vec\mE\cdot\vec\mB\ne 0$ field configurations. \\
b) In~\rs{PeriodicIm} we apply the Bogoliubov coefficient summation method~\cite{Nikishov:1979ez}, building on the result of~Ref.~\cite{Kim:2008yt} to compute the imaginary part $\mathfrak{Im}V^\mathrm{EHS}_g$ in specific field configurations. To obtain the action in the domain $|g|>2$, we must account the Landau orbitals which in exactly constant fields show behavior we are familiar with for strongly coupled $1/r$-potential in the Dirac equation, or $1/r^2$ in the Schr\"odinger equation. These singular potentials are mathematical idealizations of less singular physical forms~\cite{Case:1950an, Werner:1958zz}. We learn from past experience that a self-adjoint physical extension is required, which could be to consider localized but nearly constant fields, compare the case of a finite electric Sauter step~\cite{Evans:2022fsu}. \\ 
c) In~\rs{Ramanujan} we describe the Ramanujan periodic in $g$ formulation of the integrand entering the proper time integration, in order to obtain a unique expression for the real part of effective action based on the computed imaginary part.

In~\rs{nonpertP} we explore nonperturbative behavior of $V^\mathrm{EHS}_g$ as a function of $g$, focusing on a magnetically dominated environment:\\
a) In~\rs{CuspShape} we demonstrate the singular properties at $g=\pm 2$, in particular how the sharpness of the cusp singularity in $\mathfrak{Im}V^\mathrm{EHS}_g$ depends on EM fields in a nonperturbative manner. In magnetic dominated fields with nonvanishing $\vec\mE\cdot\vec\mB$, $\mathfrak{Im}V^\mathrm{EHS}_g$ is sharply peaked as a function of $g$ at $g=\pm 2$, and strongly suppressed for $g\neq\pm2$. \\
b) In~\rs{Stability}, we identify expansion parameter $\chi_b=\alpha b/2a$, which characterizes the onset of significant suppression of the EHS pair production result. We demonstrate that perturbative expansion in radiative order $\alpha$ corrections to $g=\pm 2$ breaks down in the $\chi_b>1$ domain. We also evaluate the effect of $\chi_b>1$ in the EM fields of magnetars. The resultant stabilizing effect dominates the otherwise monotonic enhancement of particle production by $\mB$ fields when $g=\pm 2$ exactly.\\
c) In~\rs{Savvidy}, we explore the domains of $g$ (relatively far from $g= 2$) in which asymptotic freedom arises~\cite{Rafelski:2012ui},  reproducing the results by Araujo, Napsuciale and Martinez~\cite{VaqueraAraujo:2012qa,Angeles-Martinez:2011wpn} for $|g|\le2$ and finding that the domains recur for $|g|>2$ values. We show that in these $g$ domains, $\mathfrak{Im}V^\mathrm{EHS}_g$ is essentially vanishing for magnetic dominated fields: In asymptotic freedom environment the QED vacuum state considered to one loop is practically stable. This parallels the recent finding~\cite{Savvidy:2022} in the non-Abelian QCD context where in the asymptotic free regime the vacuum stability in (chromo) magnetic dominated fields associated with the Savvidy model of the vacuum~\cite{Savvidy:1977as} was recognized.\\ 
d) In~\rs{temprep} we apply our results to extend prior work, relating $V^\mathrm{EHS}_g$ to the temperature representation. The temperature representation of $V^\mathrm{EHS}$ for electric fields~\cite{Muller:1977mm,PauchyHwang:2009rz} exhibits an inversion of spin statistics: The $g=\pm 2$ spin-1/2 ($g=0$ spin-0) action takes on a Bose (Fermi) distribution. This result was extended to $|g|\leq 2$~\cite{Labun:2012jf}, establishing a connection with the Unruh thermal background~\cite{Unruh:1976db} experienced by an accelerating observer. We extend this result to $|g|>2$, and consider the magnetic and electric field effect separately. 

In~\rs{summary2} we review our main results and discuss their implications and potential for additional study of asymptotic behavior of $V^\mathrm{EHS}_g$ for strong fields incorporating $\vec\mE\cdot\vec\mB$. We address challenges regarding convergence of perturbative QED in strong field environments. We further discuss how the singular effects we uncovered may be indicating presence of a 2nd order phase transition in magnetically dominated quasi-constant strong QED fields.

\section{EHS effective action for $|g|\leq 2$}
\label{KGPfirst}

\subsection{Proper time evaluation}
 
We summarize the proper time formulation of EHS effective action with $g\neq2$, which turns out to be limited to the domain $|g|\leq 2$. Schwinger~\cite{Schwinger:1951nm} in his manifestly covariant and gauge invariant approach employed the \lq{squared}\rq\ Dirac equation, the product of the Dirac equation with its negative mass counterpart. To incorporate anomalous magnetic moment in this approach Kruglov extended the second order fermion~\cite{Morgan:1995, Espin:2013, Espin:2015bja} wave equation to $g\neq2$, referred to as the KGP formulation~\cite{Steinmetz:2018ryf}
\begin{align}
\label{KGPgGeneral}
\Big((i\partial_\mu-eA_\mu)^2-m^2-\fracg\frac e2\sigma^{\mu\nu}F_{\mu\nu}\Big)\Psi=0
\;,
\end{align}
where
$\sigma^{\mu\nu}= \frac i2[\gamma^\mu,\gamma^\nu]$, $F_{\mu\nu}$ denotes the EM tensor, and
\begin{align}
\frac 12\sigma^{\mu\nu}F_{\mu\nu}=i\gamma^5\vec\Sigma\cdot\vec\mE-\vec\Sigma\cdot\vec\mB
\;,
\end{align}
with Pauli-Dirac matrices $\vec\Sigma=\gamma^5\gamma^0\vec\gamma$.

Kruglov used~\req{KGPgGeneral} to formulate a $g$-dependent modification to Schwinger's proper time evolution operator Eq.\,(2.33) in~\cite{Schwinger:1951nm}, with \lq Hamiltonian\rq
\begin{align}
\label{Hkgp}
H_{\mathrm{KGP}}=&\;
\Pi^2-\fracg\frac e2\sigma_{\mu\nu}F^{\mu\nu}
\;,
\end{align}
where $\Pi_\mu=p_\mu-eA_\mu$. The resulting spin $1/2$ action with $g\neq2$
\begin{align}
\label{3.22}
V^\mathrm{EHS}_g=&\,
\frac{i}{2}\int_0^\infty\frac{du}{u}e^{-im^2u}\mathrm{tr}\left<x\right|e^{-iH_\mathrm{KGP} u}\left|x\right>
\\ \nonumber
=&\;
\frac{1}{32\pi^2}\int_0^\infty\frac{du}{u^3}e^{-im^2u}
\frac{e^2u^2ab\; \mathrm{tr}\;\!e^{iu(g/2)e\sigma F/2}}{\sinh(eau)\sin(ebu)}
\;,
\end{align}
where the pre-factor follows the units of Schwinger where $\alpha=e^2/4\pi$, for a review see~\cite{Dunne:2004nc}. The electromagnetic field invariants are obtained from the eigenvalues ($\pm a,\,\pm ib$) of EM tensor $F^{\mu\nu}$:
\begin{align}
\label{invariants}
a^2-b^2=\vec\mE\,^2-\vec\mB\,^2\equiv 2S\;,\quad
a^2b^2=(\vec\mE\cdot\vec\mB\,)^2\equiv P^2
\;.
\end{align}
Eigenvalue $a$ is \lq electric-like\rq, following $a\to|\vec \mE|$ in the limit $b\to0$. Similarly the \lq magnetic-like\rq\ value $b$ follows $b\to|\vec\mB|$ for $a\to0$.  In the case of parallel electric and magnetic fields, the expressions also simplify to $a\to|\vec\mE|,\,b\to|\vec\mB|$.

Evaluation of~\req{3.22} is straightforward since only the spin-dependent trace term is affected by $g\neq2$. The resulting Kruglov~\cite{Kruglov:2001dp} action is
\begin{align}
\label{3.22b}
&V^\mathrm{EHS}_g=\frac{1}{8\pi^2}\int_0^\infty\frac{du}{u^3}e^{-i(m^2-i\epsilon)u}
F(eau,ebu,\fracg)\;,
\nonumber \\ 
&F(x,y,\fracg)
=
\frac{ x \cosh(\fracg x)}{\sinh(x)}\frac{y\cos(\fracg y)}{\sin(y)}-1
\;,\qquad\left|\fracg\right|\leq 1
\;.
\end{align}
However,~\req{3.22b} is convergent only for $|g|\leq 2$. When $|g|>2$, the proper time integration diverges  as is seen considering  
\begin{align}
\!\!\!\!\!
\lim_{u\to\infty}e^{-i(m^2-i\epsilon)u}\frac{ eau \cosh(\fracg eau)}{\sinh(eau)}\frac{ebu\cos(\fracg ebu)}{\sin(ebu)}\neq0
\;.
\end{align}
The $\cosh$ expression (numerator) is outgrowing the $\sinh$ contribution (denominator) for large $u$ for any field strength $a$. Therefore when we refer to Kruglov action we now will write $V^\mathrm{EHS}_{g\le 2}$, the equal sign recreates the original EHS effective action, which for numerical expediency is often presented after path of integration is rotated $u\to -is$. However, this rotation can only be considered for convergent integrands and thus for $|g|\le2$, but not for  $|g|>2$. Even so, we note that allowing for arbitrary field configurations and field strengths  the proper time convergence issue for $|g|>2$ is persistent and present for any choice of paths in~\req{3.22b} in the complex proper time plane. For example for $u\to -is$ the problem returns for sufficiently strong $b$-fields. The morale here is that when performing mathematically incorrect transformations we alter the nature of the spurious divergence, we cannot entirely eliminate it.

We observe further that this divergence cannot be alleviated by renormalization; the subtraction -1 in~\req{3.22b} removes the zero point energy. A second subtraction removes the charge renormalizing logarithmically divergent contribution seen for $u\to 0$. The related regularization is accomplished by the conventional subtractions including in the integrand of~\req{3.22b} 
\begin{align}
\label{renormpart}
&F(x,y,\fracg)\to F(x,y,\fracg) - \frac{x^2-y^2}6\Big(3\left(\fracg\right)^2-1\Big)
\;,
\\ \nonumber
&\left|\fracg\right|\leq 1
\,;
\end{align}
for further detail on the $g$-dependent renormalization see~\cite{Angeles-Martinez:2011wpn, VaqueraAraujo:2012qa}.

The integrand of~\req{3.22b} contains a deformation of the integration contour described by shifting poles of integrand by giving as usual a vanishingly small negative imaginary component to the mass, $m^2-i\epsilon$, according to the Feynman prescription allowing for computation of the residues:
\begin{align} 
\label{ImvactovacKrug}
\mathfrak{Im}V^\mathrm{EHS}_{g\le 2}
=&\;
\frac{e^2ab}{8\pi^2}\sum_{l=1}^\infty 
\frac{(-1)^le^{-l\pi m^2/ea}} {l}\cos(\fracg l \pi) 
\\ \nonumber
&\;\times
\frac{\cosh(\fracg l \pi b/a)}{\sinh(l\pi b/a)}
\;,\qquad\left|\fracg\right|\leq 1
\;.
\end{align}
This defines and allows computation of the imaginary part of the effective action as was described in Refs.\,\cite{Heisenberg:1935qt,Schwinger:1951nm}

\subsection{Temperature representation}

The effective action $V^\mathrm{EHS}_g$ can be reformatted into the temperature representation. We summarize the $|g|\leq 2$ extension~\cite{Labun:2012jf} of the original $g=\pm 2$ result obtained in~\cite{Muller:1977mm}. To obtain the temperature form for pure electric fields ($b\to0$),~\req{3.22b} becomes, after rotating the integration contour $u\to-is$,
\begin{align}
&V^\mathrm{EHS}_{g\le 2}\to-\frac{1}{8\pi^2}\int_0^\infty\frac{ds}{s^3}e^{-m^2s}
\Big(\frac{ eas\cos(\fracg eas)}{\sin(eas)}-1\Big)
\;.
\end{align}
We apply meromorphic expansion~\cite{Muller:1977mm, Cho:2000ei, Ramanujan}, 
\begin{align} 
\label{meromorphicE}
&\frac{ eas\cos(\fracg eas)}{\sin(eas)}-1=-\frac{e^2a^2s^2}6\Big(3\left(\fracg\right)^2-1\Big)
\\ \nonumber
&\;
+ 2e^2a^2s^4\sum_{n=1}^\infty\frac{(-1)^n\cos(\fracg n\pi)}{n^2\pi^2(s^2-n^2\pi^2/e^2a^2)}
\;,
\end{align}
and remove the first term on the RHS, which is absorbed by charge renormalization. Plugging~\req{meromorphicE} into the proper time expression~\req{3.22b} and exchanging summation with integration,
\begin{equation} 
\label{VeffE}
V^\mathrm{EHS}_{g\le 2}
= 
\frac{e^2a^2}{4\pi^2}
\sum_{n=1}^\infty \int_{0}^{\infty}ds\,s\,e^{-m^2s} 
\frac{(-1)^{n+1}\cos(\fracg n\pi)}{n^2\pi^2(s^2-n^2\pi^2/e^2a^2)}
\;.
\end{equation}
Substituting $s\to n\pi s/ea$ and exchanging summation with integration once more,
\begin{align} 
\label{VeffE2}
V^\mathrm{EHS}_{g\le 2}=&\; 
-\frac{e^2a^2}{8\pi^2}
 \int_{0}^{\infty}ds\left(\frac {1}{s-1+i\varepsilon}+\frac {1}{s+1} \right)
 \\ \nonumber
&\;\times
\sum_{n=1}^\infty e^{-n\pi m^2s/ea} 
\frac{(-1)^n\cos(\fracg n\pi)}{n^2\pi^2}
\;.
\end{align}
We introduced the path prescription, see \lq$i\varepsilon$\rq\ equivalent to the negative imaginary part $m^2\to m^2-i\varepsilon$ inherent to proper time integration; setting $\mathfrak{Im}[1/(s-1+i\varepsilon)]=-\pi i\delta(s-1)$ we find~\req{ImvactovacKrug}.

Integrating~\req{VeffE2} by parts, 
\begin{align}
\label{VeffE3}
V^\mathrm{EHS}_{g\le 2} = &
-v\int_{0}^{\infty}\!\!\! ds \ln(s^2-1+i\varepsilon) 
\\ \nonumber
&\;\times
\sum_{n=1}^\infty e^{-n\beta s} 
\frac{(-1)^n\cos\frac{g n\pi}{2}}{n\pi}\;, 
\end{align}
where\\[-1cm]
\begin{equation} 
 \beta=\frac{\pi m^2}{ea}\;,\qquad 
 v=\frac{m^4}{8\pi^3\beta} \;. 
\end{equation}
 Summing~\req{VeffE3} over $n$ we obtain 
\begin{align} 
\label{VeffE4}
V^\mathrm{EHS}_{g\le 2}
= &\;
v
\!\!\int_{0}^{\infty}\!\!\!\!\!ds \ln(s^2-1+i\varepsilon) 
\frac{1}{2}\sum_\pm \ln\left(1+e^{-\beta s}e^{\pm i\fracg \pi}\right)
\,,
\end{align}
One last manipulation of the proper time integrand in~\req{VeffE4} leads to the exponential with energy multiplied by inverse temperature, Eq.\,(12) in~\cite{Labun:2012jf}, from which we recover the $g=\pm2$ limits given by Eq.\,(7) in~\cite{Muller:1977mm}. The spectral function characterizing the density of virtual particle excitations is $\ln(s^2-1+i\varepsilon)$. The temperature representation has Bose gas character for $g=\pm 2$, and Fermion character for the spinless EHS result, here the limit at $g=0$, see table 1 of~\cite{Labun:2012jf}. Interestingly, the $g=1$ case yields the Unruh temperature~\cite{Unruh:1976db} as noted in Ref.\,\cite{Labun:2012jf}: For $g=1$ in~\req{VeffE4} the $\cos$-term disappears, hence we can redefine in the final exponential $2\beta\to \beta^\prime$, introducing the Unruh temperature in the context of a Fermi function format.

\section{Euler-Heisenberg action for arbitrary $g$, and pseudoscalar $\vec\mE\cdot\vec\mB$}
\label{pseduoNonpert}

Properties of the effective action $V^\mathrm{EHS}_g$ which are not accessible to prior efforts based on the proper time formulation can be explored using alternate methods. A complete result for $V^\mathrm{EHS}_g$ including $|g|>2$ was obtained for the case of a pure magnetic field in~\cite{Rafelski:2012ui}, employing the Weisskopf method of summing Landau energy eigen values. The result was a periodic in $g$ action. More recently, we obtained the action for all $g$ for the pure electric field case~\cite{Evans:2022fsu}, employing Nikishov's method of summing Bogoliubov coefficients~\cite{Nikishov:1979ez}. Kim, Lee and Yoon used the second order fermion KGP equation to produce a convenient single expression accounting for both $g=\pm 2$ and $g=0$ solutions~\cite{Kim:2008yt}. 

This prior work on $V^\mathrm{EHS}_g$ motivates our pursuit of the general case for arbitrary $\mE$ and $\mB$ fields. It was postulated~\cite{Rafelski:2012ui} that the effective action when both $\mE$ and $\mB$ are present exhibits similar behavior as the pure magnetic result. That is, the periodic $g$-dependence of the magnetic result applies also to the configurations with nonzero $\vec\mE\cdot\vec\mB$, producing a convergent action. However, such a behavior implies a pseudoscalar $\vec\mE\cdot\vec\mB$-dependent cusp at $g=\pm2$. Due to its singular nature, such a feature cannot be proven by analytical continuation of the previous results which consider the $\mE$ and $\mB$ cases separately. Thus it is necessary to obtain $V^\mathrm{EHS}_g$ with $\vec\mE\cdot\vec\mB$ independently, which we present below.

\subsection{Klein-Gordon-Pauli solution}
\label{KGBpseudo}

In order to use the Bogoliubov coefficient summation procedure to obtain the spin $1/2$ $g\ne 2$ effective action we generalize the solution of the \lq squared Dirac equation\rq\ -- the Klein-Gordon-Paul equation -- to arbitrary $g$, using the Weyl representation of Dirac matrices~\cite{Steinmetz:2018ryf}:
\begin{align}
\label{KGPgeneral} 
\Big(-(i\partial_\mu-eA_\mu)^2+m^2-ge\sigma(\mB- i\lambda \mE)\Big)\Psi=0
\;,
\end{align}
where $g$ in~\req{KGPgeneral} can take arbitrary values and
\begin{align}
\sigma=\pm1/2\;,\qquad \lambda =\pm1
\;.
\end{align}
$\sigma$ is the spin projection along the axis parallel to $\mE$ and $\mB$, and $\lambda$ accounts for the reduction in degrees of freedom from the 4-spinor Dirac representation to the 2-spinor Weyl form. We write $\mE=a=|\vec\mE|$ and $\mB=b=|\vec\mB|$: We work in the reference frame where both fields are parallel. Without restriction of generality fields are chosen to point in $x$-direction. 

The wave function $\Psi$ contains the Weyl 2-spinor
\begin{align}
\label{psiSplit0}
\Psi=\psi X\;,\qquad
X=
\frac1{\sqrt2}\begin{pmatrix} 1 \\ \pm1\end{pmatrix}
\;.
\end{align}
For static homogenous $\mE$ and $\mB$ pointing in the $x$-direction, the 4-potential is given by
\begin{align}
A^\mu=(0, -\mE t, 0, \mB y)
\;,
\end{align}
using the temporal gauge description of the electric field component. The KGP equation becomes  
\begin{align}
\label{KGPg}
&\;\Big\{\partial_t^2+
(p_x-e\mE t)^2-\partial_y^2
+(p_z+e\mB y)^2+m^2
\\ \nonumber
&\;\;\;
-ge\sigma(B- i\lambda \mE)\Big\}\psi=0
\;.
\end{align}

\req{KGPg} allows us to separate variables in the solution $\psi$ as:
\begin{align}
\label{psiSplit}
\psi=e^{i(p_xx+p_zz)}u(y)f(t)
\;.
\end{align}
We first solve for the $u$ component of $\psi$ that is influenced only by $\mB$. We rewrite~\req{KGPg} as
\begin{align}
\label{KGPg2}
&\;\Big(\partial_t^2+(p_x-e\mE t)^2+ig e\sigma\lambda\mE +\hat K^2\Big)\psi=0
\;,
\end{align}
where operator
\begin{align}
\hat K^2=-\partial_y^2+(p_z+e\mB y)^2+m^2-ge\sigma B
\;.
\end{align}
Introducing
\begin{align}
y=\frac\eta{\sqrt{e\mB}}-\frac{p_z}{e\mB}
\;,
\end{align}
we recognize that $\hat K^2$ provides a harmonic oscillator equation satisfying
\begin{align}
\label{Khat}
\hat K^2u=K_n^2u
\;,
\end{align}
solved by 
\begin{align}
\label{uForm}
u=H_n[\eta]\frac{(e\mB)^{1/4}e^{-\eta^2/2}}{\sqrt{2^nn!\sqrt{\pi}}}
\;.
\end{align}
$H_n$ is the Hermite polynomial describing the $n$th Landau level, and eigenvalues
\begin{align}
\label{KnForm}
 K_n^2(g,\sigma) = m^2+e\mB(2n+1-g\sigma )\;,\;\;\;
 n=0,\pm1,\pm2,\ldots
\;.
\end{align}
$n$ can have both positive and negative values. In the next section we will determine which $n$ states span the physical Hilbert space. 

We return to find $f$, the $\mE$-dependent contribution to the wavefunction $\psi$ defined in~\req{psiSplit}. Plugging~\req{Khat} and~\req{KnForm} into~\req{KGPg2}, the KGP expression reduces to the equation for an inverted harmonic oscillator potential
\begin{align}
\label{KGPg2b}
&\;\Big(\partial_t^2+(p_x-e\mE t)^2+ig e\sigma\lambda\mE + K_n^2\Big)f=0
\;.
\end{align}
We translate~\req{KGPg2b} into a parabolic cylinder differential equation by introducing the following variables
\begin{align}
\label{Zfirst}
Z=\sqrt{\frac{2}{e\mE}}e^{i\pi/4}(p_x-e\mE t)
\;,
\end{align}
and 
\begin{align}
\label{rhofirst}
\rho(g,\sigma,\lambda)
=&\;
-\frac 12+\fracg\sigma\lambda -i\frac{ K_n^2}{2e\mE}
\;,
\end{align}
which obeys the relation
\begin{align}
\label{rhoform}
\rho^*(g, \sigma,\lambda )=-\rho(g,\sigma ,-\lambda)-1
\;.
\end{align}
Plugging~\req{Zfirst} and~\req{rhofirst} into~\req{KGPg2b}, we obtain
\begin{align}
\label{KGPg2c}
&\;\Big(\partial_t^2-\frac{ie\mE}{2}Z^2+2ie\mE\Big(\rho+\frac12\Big)\Big)f=0
\;,
\end{align}
which is solved by
\begin{align}
\label{PsiSol}
f=D_\rho(Z)
\;,
\end{align}
where parabolic cylinder function $D$ has index $\rho$ given by~\req{rhofirst}.

\subsection{Bogoliubov coefficient summation}
\label{PeriodicIm}
 
The vacuum-to-vacuum amplitude in a constant applied field~\cite{Nikishov:1979ez}
\begin{align}
\label{vactovac}
\left<0 _{t=+\infty}|0_{t=-\infty}\right>=e^{iL^3TV^\mathrm{EHS}_g}
\;,
\end{align}
where $L^3T$ = volume $\times$ time.~\req{vactovac} can be written as the product of the probabilities for each negative (positive) energy state ($n$) at $t=-\infty$ to remain a negative (positive) energy state at $t=+\infty$:
\begin{align}
\label{vactovac2}
\left<0 _{t=+\infty}|0_{t=-\infty}\right>=&\;
\prod_{k}\left<0 _{k,\,t=+\infty}|0_{k,\,t=-\infty}\right>
\;,
\end{align}
where $k$ includes all spin and momentum states. Comparing~\req{vactovac} and~\req{vactovac2} we write the action as
\begin{align}
\label{vactovac3}
V^\mathrm{EHS}_g=&\;
\frac{i}{L^3T}\sum_{k}\ln c_k^*
\;,
\end{align}
labeling the Bogoliubov coefficient according to notation in~\cite{Nikishov:1979ez}:
\begin{align}
{c_k^*}^{-1}=\left<0 _{k,\,t=+\infty}|0_{k,\,t=-\infty}\right>
\;.
\end{align}

We obtain $c_k$ from the $t\to\pm\infty$ limits of the KGP solution~\req{PsiSol}. At $t\to-\infty$,~\req{PsiSol} takes on the form~\cite{Kim:2008yt}
\begin{align}
D_\rho(Z)_{t\to-\infty}=e^{-Z^2/4}Z^\rho
\;,
\end{align}
and at $t\to+\infty$
\begin{align}
\label{tinfty}
D_\rho(Z)_{t\to+\infty}=&\;e^{-i\pi p}D_\rho(-Z)_{t\to-\infty}
\\ \nonumber 
&\;
+\frac{\sqrt{2\pi}}{\Gamma(-\rho)}e^{-i\pi (\rho+1)/2}D_{-\rho-1}(-Z)_{t\to-\infty}
\;.
\end{align}
The coefficient of the first term in~\req{tinfty} corresponds to the tunneling amplitude through the mass gap, while the second term gives Bogoliubov coefficient 
\begin{align}
\label{cform}
c_k=\frac{\sqrt{2\pi}}{\Gamma(-\rho)}e^{-i\pi (\rho+1)/2}
\;.
\end{align}

Plugging~\req{cform} into~\req{vactovac3} we obtain the imaginary part of effective action
\begin{align}
\label{Imvactovac}
2\mathfrak{Im}V^\mathrm{EHS}_g=&\;
\frac1{L^3T}\sum_{k}\ln \Big[ \Big|
\frac{\sqrt{2\pi}}{\Gamma(-\rho)}e^{-i\pi (\rho+1)/2}
\Big|^2\Big]
\;,
\end{align}
with summation
\begin{align}
\label{sumform}
\sum_{k}=&\;\frac12\sum_{n,\sigma,\lambda} L^2 \int \frac{dp_x dp_z}{(2\pi)^2}
=
L^3T\frac{e^2\mE\mB}{8\pi^2}\sum_{n,\sigma,\lambda}
\;,
\end{align}
where the $1/2$ factor averages the $\lambda=\pm1$ contributions, and $\int dp_z=e\mB L$ and $\int dp_x=e\mE T$, Eq.\,(3.7) of~\cite{Nikishov:1979ez}. The imaginary action per unit time and volume~\req{Imvactovac} becomes
\begin{align}
\label{Imvactovac2}
\mathfrak{Im}V^\mathrm{EHS}_g=&\;
\frac{e^2\mE\mB}{16\pi^2}\sum_{n,\sigma,\lambda}
\ln \Big[ \Big|
\frac{\sqrt{2\pi}e^{-i\pi (\rho+1)/2}}{\Gamma(-\rho)}
\Big|^2\Big]
\;.
\end{align}
Carrying out summation over $\lambda$ first, we have
\begin{align}
\label{Imvactovac3}
\mathfrak{Im}V^\mathrm{EHS}_g
=&\;
\frac{e^2\mE\mB}{16\pi^2}\sum_{n,\sigma}
\ln \Big\{
\frac{2\pi e^{-i\pi (\rho(g,\sigma,1)+\rho(g,\sigma,-1))/2} }
{\Gamma(-\rho(g,\sigma,1))\Gamma(-\rho(g,\sigma,-1))}
 \nonumber \\
&\;
\!\!\!\times
\frac{2\pi e^{i\pi (\rho^*(g,\sigma,1)+\rho^*(g,\sigma,-1))/2}}
{\Gamma(-\rho^*(g,\sigma,1))\Gamma(-\rho^*(g,\sigma,-1))}
\Big\}
\;.
\end{align}
The complex conjugated terms in~\req{Imvactovac3} can be rewritten using the relation~\req{rhoform}:
\begin{align}
\label{Imvactovac4}
\mathfrak{Im}V^\mathrm{EHS}_g
=&\;
\frac{e^2\mE\mB}{16\pi^2}\sum_{n,\sigma}
\ln \Big\{
\frac{2\pi ie^{-i\pi \rho(g,\sigma,1)}}{\Gamma(-\rho(g,\sigma,1))\Gamma(1+\rho(g,\sigma,1))}
 \nonumber \\
&\;
\!\!\!\times
\frac{2\pi i e^{-i\pi \rho(g,\sigma,-1)} }
{\Gamma(-\rho(g,\sigma,-1))\Gamma(1+\rho(g,\sigma,-1))}
\Big\}
\;,
\end{align}
which then allows for use of the Euler reflection formula {\it e.g.}
\begin{align}
\frac{1}{\Gamma(-\rho(g,\sigma,\pm1))\Gamma(1+\rho(g,\sigma,\pm1))}
=-\frac{\sin(\rho(g,\sigma,\pm1)}\pi
\;.
\end{align}
This allows us to rewrite~\req{Imvactovac4} as 
\begin{align}
\label{Imvactovac5}
\mathfrak{Im}V^\mathrm{EHS}_g
=&\;
\frac{e^2\mE\mB}{16\pi^2}\sum_{n,\sigma}
\Big\{ \ln [ 1-e^{-2i\pi \rho(g,\sigma,1)}]
\\ \nonumber 
&\;\qquad\qquad\!
+\ln [1-e^{-2i\pi \rho(g,\sigma,-1)}] \Big\}
\;.
\end{align}
Using the series representation of the logarithms 
\begin{align} 
\ln [ 1-e^{-2i\pi \rho(g,\sigma,1)}]
=\sum_{l=1}^\infty \frac{e^{-2il\pi\rho(g,\sigma,1)}}l
\;,
\end{align}
and applying definition of $\rho$ in~\req{rhofirst}, 
\req{Imvactovac5} becomes
\begin{align} 
\label{Imvactovac6}
\mathfrak{Im}V^\mathrm{EHS}_g
=&\;
\frac{e^2\mE\mB}{8\pi^2}\sum_{l=1}^\infty \frac{(-1)^l e^{-l\pi m^2/e\mE}}{l}\cos(\fracg l\pi)
\\ \nonumber 
&\;\times
\sum_{n,\sigma} e^{-l\pi ( K_n^2(g,\sigma)-m^2)/e\mE}
\;,
\end{align} 
where $K_n$ is given by~\req{KnForm}, and 
\begin{align}
\label{spectrumtilde}
\!\!\frac{K_n^2-m^2}{e\mE}=\frac{\mB}{\mE}(2n+1 - g\sigma)\;,\;\;\;
n=0,\pm 1,\pm2\ldots
\;.
\end{align}

In~\rf{fig:Espectrum} we plot the exponential terms $e^{-l\pi ( K_n^2-m^2)/e\mE}$,~\req{Imvactovac6}, for different $n$ levels. We identify which $n$ comprise the physical spectrum, that satisfy the correct boundary conditions and preserve unitary. These are the levels for which $|e^{-l\pi ( K_n^2-m^2)/e\mE}|<1$ in~\rf{fig:Espectrum}, corresponding to the states $K_n^2\geq m^2$. For these states, the imaginary part of $V^\mathrm{EHS}_g$ vanishes as $\mE\to0$, ensuring a stable QED vacuum in absence of external fields.

The unphysical, non-normalizable $n$ values appear in~\rf{fig:Espectrum} as the cases where $|e^{-l\pi(K_n^2-m^2)/e\mE}|>1$ due to $ K_n^2<m^2$. In sufficiently strong $\mB$ fields, the Hamiltonian acquires complex eigenvalues ($K_n$ becomes imaginary) and self-adjointness is lost, causing the imaginary part of action to be nonzero even in the limit $\mE\to0$. Such ill-defined spurious states normally originate in idealized unphysical field shapes  {\it e.g.} generated by singular (Coulomb-like) potentials~\cite{Case:1950an, Werner:1958zz}. A self-adjoint extension of the Hamiltonian removes these spurious states. In our case this is the infinite extent constant fields. We verify below that such spurious states do not exist when using localized fields that are physically realizable in nature rather than constant fields.

We count the states in~\rf{fig:Espectrum} that make up the physical spectrum. For the domain $-2\leq g\leq 2$ including the well known $g=2$ case, we admit $n\ge 0$. The situation changes for $|g|>2$, where a shift in $g$ by multiples of 4 produces the corresponding change in $n$:
\begin{align}
\label{kapForm}
&e^{-l\pi ( K_n^2(g+4k,\sigma)-m^2)/e\mE}
=e^{-l\pi ( K_{n-2\sigma k}^2(g,\sigma)-m^2)/e\mE}\;,
 \nonumber \\
&\;
k=0,\pm 1, \pm 2, \ldots
\;.
\end{align}
The periodic values of $g=2+ 4k$ are crossing points, at which one state disappears from the spectrum, while another state with opposite spin projection joins the physical spectrum. Thus with each shift in $g\to g+4k$ there is a duplication of the physical spectrum. Which $n$ levels are admitted depends on the branch $-2+4k<g<2+4k$ in which $g$ resides, table~\ref{kappa+shift}.

\begin{figure}
\centering
\includegraphics[width=0.99\columnwidth]{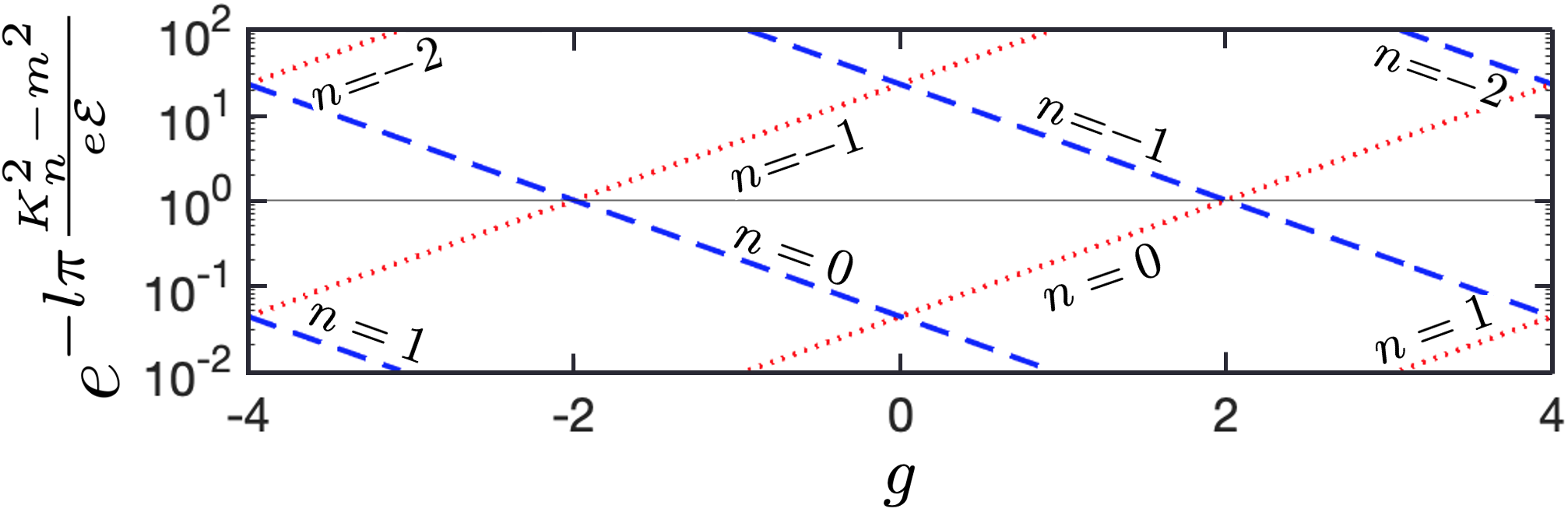}
\caption{The periodic spectrum $e^{-l\pi ( K_n^2-m^2)/e\mE}$ summed in~\req{Imvactovac6}, for the case $l=1$ and $\mB/\mE=1$. Dotted and dashed lines correspond to spin $\sigma=+1/2$ and $\sigma=-1/2$, respectively. \label{fig:Espectrum}}
\end{figure}

We note that while we determined the states in table~\ref{kappa+shift} using~\rf{fig:Espectrum} for the specific example of $l=1$ and $\mB/\mE=1$, the argument can be generalized to arbitrary field strengths. The relative strengths of $\mB$ and $\mE$ do not affect the condition $K_n^2\geq m^2$ in order for the states to be physical. The spectrum in table~\ref{kappa+shift} agrees with the result from prior work on the magnetic Weisskopf action for $|g|>2$~\cite{Rafelski:2012ui}.

\begin{table}[h]
\centering
\begin{tabular}{cc|c|c}
\cline{3-3}
& & \multicolumn{1}{ c| }{Physical spectrum} \\ 
\cline{1-3}
\multicolumn{1}{ |c }{\multirow{2}{*}{$-6<g<-2$} } & 
\multicolumn{1}{ |c| }{$\sigma=+1/2$} & $\quad\!n=-1,0,1,\ldots$ & \\ \cline{2-3}
\multicolumn{1}{ |c }{} &
\multicolumn{1}{ |c| }{$\sigma=-1/2$} & $ n=1,2,3,\ldots$ & \\ \cline{1-3}
\multicolumn{1}{ |c }{\multirow{2}{*}{$-2<g<2$} } & 
\multicolumn{1}{ |c| }{$\sigma=+1/2$} & $n=0,1,2,\ldots$ & \\ \cline{2-3}
\multicolumn{1}{ |c }{} &
\multicolumn{1}{ |c| }{$\sigma=-1/2$} & $n=0,1,2,\ldots$ & \\ \cline{1-3}
\multicolumn{1}{ |c }{\multirow{2}{*}{$2<g<6$} } & 
\multicolumn{1}{ |c| }{$\sigma=+1/2$} & $n=1,2,3,\ldots$ & \\ \cline{2-3}
\multicolumn{1}{ |c }{} &
\multicolumn{1}{ |c| }{$\sigma=-1/2$} & $\quad\! n=-1,0,1,\ldots$ & \\ \cline{1-3}
\hline \hline 
\multicolumn{1}{ |c }{$-2\!+\!4k\!<\!g\!<\!2\!+\!4k$ } & 
\multicolumn{1}{ |c| }{$\sigma$ 
} & $ n=2\sigma k,1+2\sigma k,2+2\sigma k,\ldots$ & \\ \cline{1-3}
\end{tabular}
\caption{
$n$ states for which $|e^{-l\pi ( K_n^2-m^2)/e\mE}|<1$ in~\rf{fig:Espectrum},~\req{spectrumtilde}.
\label{kappa+shift}}
\end{table}

We carry out the summation over $n$ and $\sigma$ in~\req{Imvactovac6}, applying the physical spectrum $n$ in table~\ref{kappa+shift} and~\req{kapForm}:
\begin{align} 
\label{LandauSum0}
&\sum_{\sigma=\pm\frac12}\sum_{n=2\sigma k}^\infty e^{-l\pi ( K_n^2(g,\sigma)-m^2)/e\mE}
\\ \nonumber 
&\;
=
\sum_{\sigma=\pm\frac12}\sum_{n=0}^\infty e^{-l\pi ( K_{n}^2(g+4k,\sigma)-m^2)/e\mE}
\;,
\end{align}
periodic in $g$ by shifts in $4k$. For arbitrary $|g|>2$, we choose $k$ such that we can define a periodically reset $g_k$ which lies in the principal domain 
\begin{align} 
\label{gkreset}
-2\leq g_k=g+4k\leq 2
\;.
\end{align}
Thus any summation~\req{LandauSum0} carried out for a $g$ value in the domain $|g|>2$ has an equivalent summation using $|g_k|\leq 2$. This allows us to convert the effective action with $|g|>2$ to an equivalent expression with $|g_k|\leq 2$ that is valid in the proper time formulation {\it e.g.} the perturbative electron $g$-factor resets according to
\begin{align} 
\label{gElectronreset}
g_{\mathrm{electron}}=2+\alpha/\pi \;\to\;
g_k=g_{\mathrm{electron}}-4=-2+\alpha/\pi
\;.
\end{align}
The $g_{\mathrm{electron}}>2$ can now be applied to the proper time formulation of $\mathfrak{Im}V^\mathrm{EHS}_g$,~\req{ImvactovacKrug} previously limited to $|g|\leq 2$, by resetting $g_{\mathrm{electron}}$ to the $|g_k|\leq 2$ domain.

We can now sum~\req{LandauSum0} using the series 
\begin{align} 
\sum_{n=0}^\infty e^{-l\pi\mB(2n+1)/\mE}
=&\;
\frac{1}{2\sinh(l\pi \mB/\mE)}
\;,
\end{align}
to obtain 
\begin{align} 
\label{LandauSum}
\sum_{\sigma=\pm\frac12}\sum_{n=0}^\infty e^{-l\pi ( K_n^2(g+4k,\sigma)-m^2)/e\mE}
=&\;
\frac{\sinh(\frac{g_k}2l\pi \mB/\mE)}{\sinh(l\pi \mB/\mE)}
\;.
\end{align}
We evaluate the imaginary part of $V^\mathrm{EHS}_g$ by plugging~\req{LandauSum} into~\req{Imvactovac6} and recognizing periodicity of the remaining $g$-dependent term 
\begin{align} 
\label{cosGk}
\cos(\fracg l\pi)=\cos(\fracg l\pi+2\pi k)=\cos(\frac{g+4k}{2}l\pi)
\;,
\end{align}
to obtain 
\begin{align} 
\label{Imvactovac8}
\mathfrak{Im}V^\mathrm{EHS}_g
=&\;
\frac{e^2ab}{8\pi^2}\sum_{l=1}^\infty 
\frac{(-1)^l} {l}\cos(\frac{g_k}{2}l\pi) 
\\ \nonumber 
&\;\times
e^{-l\pi m^2/ea}\frac{\cosh(\frac{g_k}{2}l\pi b/a)}{\sinh(l\pi b/a)}
\;.
\end{align}
In consideration of Lorentz invariance we wrote $\mE$ and $\mB$ in terms of the EM field tensor eigenvalues $\mE\to a$ and $\mB\to b$,~\req{invariants}.~\req{Imvactovac8} is equivalent to~\req{ImvactovacKrug} in the $|g|\leq 2$ domain. However, now instead of $g$ due to the periodic behavior the reset value $g_k$ appears,~\req{gkreset}.

As a verification of our approach we consider the $\mB\to0$ limit:~\req{Imvactovac8} becomes 
\begin{align} 
\label{ImvactovacE}
\mathfrak{Im}V^\mathrm{EHS}_g
\to&\;
\frac{e^2\mE^2}{8\pi^2}\sum_{l=1}^\infty 
\frac{(-1)^l} {l^2\pi}\cos(\frac{g_k}{2}l\pi)e^{-l\pi m^2/e\mE}
\;.
\end{align}
The limiting form \req{ImvactovacE} allows us to verify the self-adjoint extension we applied to select physical states $K_n^2\geq m^2$, the (discrete) Landau $n$ states listed in table~\ref{kappa+shift}. We recover the same self-adjoint formulation by applying finite sized fields, thus departing from the idealized (constant) fields. A localized field configuration was explored using the finite electric Sauter step~\cite{Evans:2022fsu}, which permits well-defined asymptotic states. Summing such states in the (continuum) Bogoliubov coefficient summation, no discussion is needed as to which states to retain or not, and the same periodicity in $g$ arises.

We also recover the known $g=\pm 2$ and $g=0$ limits of the imaginary part of action in table~\ref{g02}, compare Nikishov's~\cite{Nikishov:1979ez} Eq.\,(3.11) and Eq.\,(3.8), noting that the $g=0$ case differs from the spin-0 result by a factor $-2$ accounting for spin multiplicity and an extra sign accompanying loop Fermionic corrections. 

\begin{table}[th]
\centering
 \begin{tabular}{ | l | c | c | }
 \hline 
 & $g=0$ & $g=\pm 2$ \\ \hline 
 $\cos(\frac{g_k}{2}l\pi)$ & 1 & $(-1)^l$ \\ \hline
 $\cosh(\frac{g_k}{2}l\pi \mB/\mE)$ & 1 & $\cosh(l\pi \mB/\mE)$ \\
 \hline
 \end{tabular}
\caption{The $g\to0$ and $g\to \pm 2$ limits of the $g$-dependent terms in~\req{Imvactovac8}. 
\label{g02}}
\end{table}

\subsection{Ramanujan equation}
\label{Ramanujan}

Our next objective is to uniquely determine the full effective action $V^\mathrm{EHS}_g$, including the real part. For analytical functions this only requires that we know the imaginary part: Using $\mathfrak{Im}V^\mathrm{EHS}_g$ obtained in~\rs{PeriodicIm} we thus reconstruct the real part. Since we already know that our result produces a singular function of $g$ with cusps we proceed to show that there is a unique proper time integral representation of $V^\mathrm{EHS}_g$, in which the integrand in proper time representation has the pole structure required to produce the imaginary part~\req{Imvactovac8}, thereby determining $V^\mathrm{EHS}_g$. 

Our results will arise from a $g$-dependent generalization of the Ramanujan expression for meromorphic expansion of $V^\mathrm{EHS}_g$ as was proposed in Ref.\,\cite{Rafelski:2012ui}. The difference to prior work is that we have obtained $\mathfrak{Im}V^\mathrm{EHS}_g$, the imaginary part of Euler-Heisenberg-Schwinger (EHS) type effective action by explicit evaluation in~\rs{PeriodicIm}.

The meromorphic expansion of the poles of the proper time integrand is well known for $g=\pm 2$~\cite{Muller:1977mm,Ramanujan,Cho:2000ei}. We obtain an extension of the $g=\pm 2$ expression Eq.\,(6) of the work by Cho and Pak~\cite{Cho:2000ei}
\begin{align} 
\label{meromorphicG}
&F(x,y,\fracg)
=
\frac{ x \cosh(\fracg x)}{\sinh(x)}\frac{y\cos(\fracg y)}{\sin(y)}-1
 \\ \nonumber
&=
2 \sum_{n =1}^\infty\frac{(-1)^{n}\cos(\fracg n\pi)}{n\pi}
\Big\{
\frac{x^2-y^2}{n\pi}
+\frac{xy^3}{y^2-n^2\pi^2} \frac{\cosh(\fracg \frac{n \pi x}y)}{\sinh(\frac{n \pi x}y)}
\\ \nonumber 
&\;\;\;\;
-\frac{x^3y}{x^2+n^2\pi^2}\frac{\cosh(\fracg \frac{n \pi y}x)}{\sinh(\frac{n \pi y}x)}
\Big\}
\;,\qquad\quad
\left|\fracg\right|\leq 1 \;.
\end{align}
We apply the following Fourier series which is defined for $|g|\leq 2$ and periodic otherwise when $|g|>2$:
\begin{align} 
\label{FourierCosh}
&\frac{\cosh(\fracg \pi z)}{\sinh(\pi z)}=a_0+2\sum_{\ell =1}^\infty a_\ell\cos(\fracg \ell\pi)
\;,
\\ \nonumber 
&\;
a_\ell= \frac12\int_0^2 dg\cosh(\fracg \pi z)\cos(\fracg \ell\pi) 
= \frac{(-1)^\ell }{\pi z(1+\ell^2/z^2)}
\;.
\end{align}
We plug~\req{FourierCosh} into~\req{meromorphicG} to obtain
\begin{widetext}
\begin{align}
\label{meroexpB0} 
F(x,y,\fracg)
=&\;2\sum_{n =1}^\infty\frac{(-1)^{n}\cos(\fracg n\pi)}{n^2\pi^2}
\Big(x^2-y^2+\frac{y^4}{y^2-n^2\pi^2} - \frac{x^4}{x^2+n^2\pi^2}\Big)
\nonumber \\ 
&\!\!\!\!\!\!\!\!\!\!\!\!\!\!\!\!\!\!\!\!\!\!\!\!\!\!\!\!\!\!\!\!\!\!
+ 4\!\!\sum_{n,\ell =1}^\infty\frac{(-1)^{n+\ell }\cos(\fracg n\pi)\cos(\fracg \ell \pi)}{\pi^2}
\Big(
\frac{y^4x^2}{(y^2-n^2\pi^2)(n^2x^2+\ell ^2y^2)}
- \frac{x^4y^2}{(x^2+n^2\pi^2)(n^2y^2+\ell ^2x^2)}
\Big)
\;.
\end{align}
In the last term in parenthesis of the second line~\req{meroexpB0} we exchange indices $n\leftrightarrow \ell$ to obtain 
\begin{align}
\label{meroexpB} 
\!\!\!\!\!\!
F(x,y,\fracg)=
2\sum_{n =1}^\infty\!
\frac{(-1)^{n}\cos(\fracg n\pi)}{n^2\pi^2}
\Big(x^2-y^2+\frac{y^4}{y^2-n^2\pi^2} - \frac{x^4}{x^2+n^2\pi^2}\Big)
+ 4x^2y^2\!\!\sum_{n,\ell =1}^\infty\!\!
\frac{(-1)^{n+\ell }\cos(\fracg n\pi)\cos(\fracg \ell \pi)}{(y^2-n^2\pi^2)(x^2+\ell ^2\pi^2)}
\,.
\end{align}
\end{widetext}
Equation\,(\ref{meroexpB}) reproduces Kruglov's result~\req{3.22b} for $|g|\leq 2$, but also happens to be periodic in $g$ for values greater than 2: All $g$-dependence in~\req{meroexpB} is in the $\cos$ terms, which are periodic according to~\req{cosGk}. Thus from~\req{meroexpB} it follows that we may simply replace in Kruglov's result~\req{3.22b}
\begin{align} 
\label{meroexpBper}
F(eau,ebu,\fracg)=F(eau,ebu,\frac {g_k}2)\;,
\quad
k=0,\pm1,\pm2,\ldots 
\;,
\end{align}
where we convert $F$ to its equivalent expression in terms of $|g_k|\leq 2$,~\req{gkreset}. We thus have shown that~\req{meroexpBper} can be used in the proper time integration, where any $|g|>2$ is expressed in terms of the proper time integrand within the $|g_k|\leq 2$ domain. 
 
We verify that the expression $F(eau,ebu,\frac{g_k}2)$ for arbitrary $g$ contains the correct pole structure to produce the imaginary part of action we obtained in~\rs{PeriodicIm}. We plug~\req{meroexpB} into the proper time integral~\req{3.22b} and compute residues of the poles to obtain
\begin{align} 
\label{meroIm}
&\mathfrak{Im}V^\mathrm{EHS}_g 
= \frac{ e^2a^2}{8\pi^3} 
\sum_{n=1}^\infty \frac{(-1)^n\cos(\frac {g_k}2 n\pi)} {n^2}
\\ \nonumber
&\!\times
\Big(
e^{-n\pi m^2/ea}
+\! 2(e b n)^2\sum_{\ell=1}^\infty
\frac{(-1)^{\ell }\cos(\frac {g_k}2 \ell \pi) e^{-\ell \pi m^2/ea}} {e^2a^2n^2+e^2b^2\ell ^2}
\Big)
\;.
\end{align}
Swapping indices $n\leftrightarrow\ell$ in the second term we obtain
\begin{align} 
\label{meroImFinal}
\mathfrak{Im}V^\mathrm{EHS}_g 
=&\; \frac{ e^2a^2}{8\pi^3} 
\sum_{n=1}^\infty \frac{(-1)^ne^{-n\pi m^2/ea}} {n^2}\cos(\frac {g_k}2 n\pi)
\\ \nonumber
&\;\times
\Big( 1 + 2\sum_{\ell=1}^\infty\frac{(-1)^{\ell }\cos(\frac {g_k}2 \ell \pi) } {1+a^2\ell^2/b^2n^2} \Big)
\,.
\end{align}
Recognizing the Fourier transform~\req{meromorphicG}, we recover~\req{Imvactovac8}.

Having verified the periodic in $g$ proper time integral representation~\req{meroexpBper}, we obtain the $|g|>2$ extension of~\req{3.22b}:
\begin{align}
\label{3.22c}
&V^\mathrm{EHS}_g=\frac{1}{8\pi^2}\int_0^\infty\frac{du}{u^3}e^{-i(m^2-i\epsilon)u}
F(eau,ebu,\frac{g_k}2)\;,
 \nonumber \\ 
&F(eau,ebu,\frac{g_k}2)
=
\frac{ eau \cosh(\frac{g_k}2 eau)}{\sinh(eau)}\frac{ebu\cos(\frac{g_k}2 ebu)}{\sin(ebu)}-1
\;.
\end{align}
\req{3.22c} ensures that $|g|>2$ action has an equivalent form using $|g_k|\leq 2$ in~\req{gkreset} producing a result for all $g$ with the proper time integration.

\section{Singular and nonperturbative properties}
\label{nonpertP}

This section describes several properties of $V^\mathrm{EHS}_g$ that rely on the nonperturbative treatment of the magnetic moment. Without doubt other interesting properties will appear in the future; this should be considered an initial consideration of the opportunities opened up by applying a nonperturbative method to achieve resummation of diagrams, see~\rf{fig:sum}. The highly singular outcome of the magnetic moment anomaly should be also a warning not to draw quantitative conclusions too soon about physical processes based on a relatively small subset of all diagrams the original EHS effective action $V^\mathrm{EHS}=V^\mathrm{EHS}_g|_{g=\pm2}$ represents.

\subsection{Singular properties of $V^\mathrm{EHS}_g$ as a function of $g$}
\label{CuspShape}
As a consequence of the periodicity in $g$, the effective action $\mathfrak{Im}V^\mathrm{EHS}_g$ peaks at the singular points $g=\pm 2$. This is particularly well visible in considering the imaginary part: Normalizing~\req{Imvactovac8} to the $g=\pm 2$ EHS value
\begin{align} 
\label{ImvactovacR}
R(g)=&\; \;\frac{\mathfrak{Im}V^\mathrm{EHS}_g}{\mathfrak{Im}V^\mathrm{EHS}} 
\\ \nonumber
=&\; 
\displaystyle\frac{
\displaystyle\sum_{l=1}^\infty \!
\displaystyle\frac{(-1)^le^{-l\pi m^2/ea}} {l}\!\cos(\frac{g_k}{2}l\pi) 
\frac{\cosh((g_k/2)l\pi b/a)}{\sinh(l\pi b/a)}
}
{\displaystyle\sum_{l=1}^\infty \frac{e^{-l\pi m^2/ea}} {l} 
\frac{\cosh(l\pi b/a)}{\sinh(l\pi b/a)}}
,
\end{align}
where we use~\req{Imvactovac8} both for $g=2$ in denominator and as a general expression in the numerator. Considering the leading terms in both denominator and numerator we note that $\frac{\cosh((g_k/2)\pi b/a)}{\sinh(\pi b/a)}$ in the numerator is always equal or smaller than $\frac{\cosh(\pi b/a)}{\sinh(\pi b/a)}=\coth(\pi b/a)$ in the denominator, since we converted $\mathfrak{Im}V^\mathrm{EHS}_g$ for arbitrary $|g|>2$ to an equivalent expression in terms of periodically reset $|g_k|\leq 2$,~\req{gkreset}. As a result, $\mathfrak{Im}V^\mathrm{EHS}_g$ is at normalized maximum at $g_k=2$, corresponding to values $g=2+4k$, $k=0,\pm1,\pm2,\ldots$. 

It is of considerable interest to understand analytically how $g\neq2$ suppresses $\mathfrak{Im}V^\mathrm{EHS}_g$. To show this we use the $\cosh$ addition theorem
\begin{align} \label{CoshAdd}
\displaystyle\frac{\cosh(\frac{g_k}2l\pi b/a)}{\sinh(l\pi b/a)}
=&\;\cosh((1-|g_k|/2)l\pi b/a)\coth(l\pi b/a) 
\nonumber \\ 
&\;
-\sinh((1-|g_k|/2)l\pi b/a)\;,
\;
\end{align}
where $(1-|g_k|/2)\geq 0$. Using in~\req{ImvactovacR} relation~\req{CoshAdd} with $\coth(l\pi b/a)\to\coth(l\pi b/a)+ 1-1$ we obtain after some algebra 
\begin{widetext}
\begin{align} 
\label{Rlim}
R(g)=&\;
\frac{
\displaystyle\sum_{l=1}^\infty 
\frac{(-1)^le^{-l\pi m^2/ea}} {l}\cos(\frac{g_k}{2}l\pi) 
\Big(e^{-(1-|g_k|/2)l\pi b/a} + e^{-l\pi b/a}
\frac{\cosh((1-|g_k|/2)l\pi b/a)}{\sinh(l\pi b/a)}\Big) 
}
{\displaystyle\sum_{l=1}^\infty \frac{e^{-l\pi m^2/ea}} {l} \coth(l\pi b/a)}
\;.
\end{align} 
\end{widetext}
Remembering $1+e^{-x}/\sinh x=\coth x$ we recover $R\to 1$ for any value of $b$ for $g_k\to \pm 2$.

We find suppression of the imaginary part of action in the presence of $(1-\frac{|g_k|}2) b/a\gg 1$ for all field strengths and $g$ values. The larger $(1-\frac{|g_k|}2) b/a$, {\it i.e.} the more magnetic-interaction dominated is the particle-EM field configuration, the more pronounced is the resultant suppression of the imaginary part of the action due to the first term in parenthesis in~\req{Rlim} which dominates, causing the magnitude of $R(g)$ to drop exponentially. We further note that this behavior proves the conjecture from our prior work~\cite{Evans:2018kor} in which the suppression for $g\neq2$ was associated with an equivalent increase of mass:
\begin{align} 
\label{mshift}
m^2\to m^2 +(1-\frac{|g_k|}2) eb
\;.
\end{align}

In quantitative evaluation we consider the behavior of $R(g)$ in terms of dimensionless EM field quantities
\begin{align} 
\label{abdef2}
\tilde a=\frac{ea}{m^2}=\frac{a}{E_\mathrm{cr}}\;,\qquad 
\tilde b=\frac{eb}{m^2}=\frac{b}{B_\mathrm{cr}}
\;,
\end{align} 
that is in units of the EHS critical fields, in SI-units format
\begin{align} 
\label{EBcritical}
 E_\mathrm{cr}&=\frac{m^2}{e} \to \frac{m^2c^3}{e\hbar}=1.3233\times 10^{18} \,\mathrm{V/m}
 \;,
 \nonumber \\
 B_\mathrm{cr}&=\frac{m^2}{e} \to \frac{m^2c^2}{e\hbar}=\frac{1.3233\times 10^{18}}{2.9979\times 10^8}= 4.4140\times 10^9\,\mathrm{T}
 \;.
\end{align} 
The numerical values in~\req{EBcritical} are obtained cancelling in one $mc^2=0.51100$\,MeV the \lq$e$\rq\ and using Compton wavelength $\hbar/mc=386.16\times 10^{-15}$\,m.

In~\rf{fig:KnE} we show $R$ from~\req{Rlim} for three field configurations. The pure electric field case $\tilde a=1$ with $\tilde b=0$ (blue, dashed line) exhibits a smooth nonsingular $g$-dependence, in agreement with prior work on the pure electric case where action is differentiable at $g=2$~\cite{Evans:2022fsu}. 

A different result emerges for field configurations where both electric and magnetic fields are nonzero: We have cusps at reset points $g_k$, in~\rf{fig:KnE} seen at $g=\pm 2$. For the $\tilde a=\tilde b=1$ configuration (red, dot-dashed line) case, the cusp is clearly visible. The cusp becomes more sharply peaked with increasing magnetic field ($\tilde a=1$ and $\tilde b=10$, black, solid).

To demonstrate that the cusp points, here as example $g= 2$, in~\rf{fig:KnE} are truly singular for $b\ne 0$ (and not only a sharply peaked function that is smooth at $g=2$), we compute the discontinuity in the derivative of $\mathfrak{Im}V^\mathrm{EHS}_g$ with respect to $g$. Differentiating~\req{Imvactovac8} with respect to $g$ at opposite sides of $g=2$ and taking the difference we obtain 
\begin{align} 
\label{Imvactovac9}
&\frac{\partial \mathfrak{Im}V^\mathrm{EHS}_g}{\partial g}\Big|_{g=2_+}
\!\!\!-\frac{\partial \mathfrak{Im}V^\mathrm{EHS}_g}{\partial g}\Big|_{g=2_-}
\nonumber \\ 
=&\;
\frac{e^2ab}{8\pi^2}\sum_{l=1}^\infty 
\frac{(-1)^le^{-l\pi m^2/ea}} {l}\cos(\frac{g}2 l\pi)\Big|_{g=2}
\\ \nonumber
&\;\times
\frac{\partial}{\partial g}
\frac{\cosh(\frac{g-4}2 l\pi b/a)-\cosh(\fracg l\pi b/a)}{\sinh(l\pi b/a)}
\Big|_{g=2}
\nonumber \\ 
=&\;
\frac{e^2b^2}{8\pi}\sum_{l=1}^\infty 
e^{-l\pi m^2/ea}
=\frac{e^2b^2}{8\pi(e^{\pi m^2/ea}-1)}
\;,
\end{align}
where $g=2_+$ and $g=2_-$ denote $g$ approaching 2 from opposite sides of the cusp. $2_-$ is already smaller than 2 and thus requires no $g$-reset, while $2_+$ is barely above 2 thus requiring reset $2_+\to 2_+-4$ according to~\req{gkreset}. We see that only at exactly $b=0$ the discontinuity vanishes.

\begin{figure}
\centering
\includegraphics[width=0.95\columnwidth]{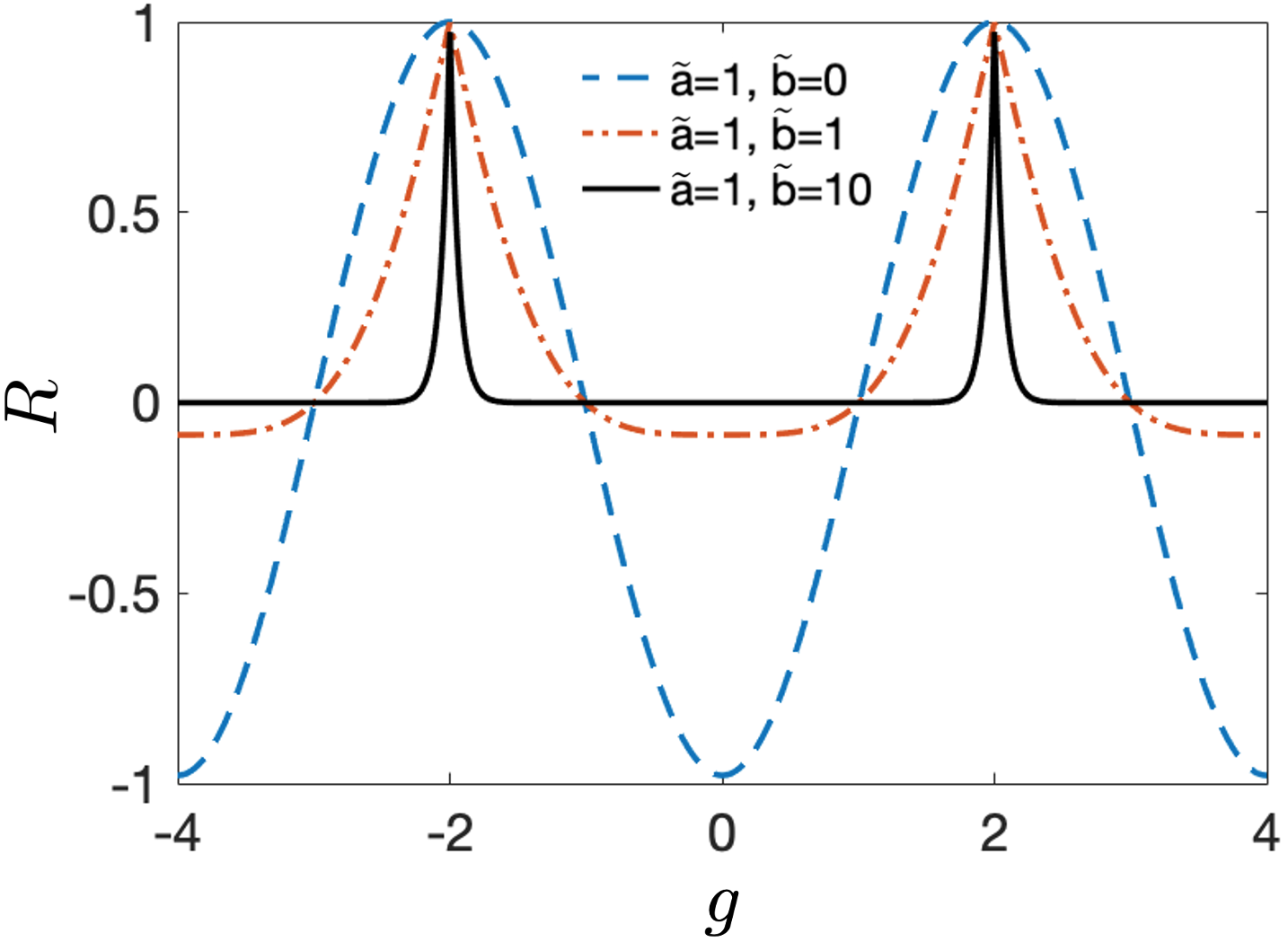} 
\caption{The normalized $\mathfrak{Im}V^\mathrm{EHS}_g$ expression $R$,~\req{Rlim}, plotted as a function of $g$ for $\tilde a=1$ and three different $\tilde b=0, 1, 10$ magnetic fields,~\req{abdef2}.
\label{fig:KnE} }
\end{figure}

\subsection{Relevance of non-perturbative treatment of\\ electron magnetic moment}
\label{Stability} 

When $g$ is not exactly equal to 2,~\rf{fig:KnE} demonstrates a stabilizing effect on the vacuum. In magnetic dominated fields $b/a\to\infty$, the width of the peak in $\mathfrak{Im}V^\mathrm{EHS}_g$ shrinks until the particle production can only occur at points $g=2+4k$, $k=0,\pm1,\pm2,\ldots$. Thus even a small deviation from $g=\pm 2$ can cause $\mathfrak{Im}V^\mathrm{EHS}_g$ to fall off the peak (and tend to zero), suppressing particle production.

The above indicates that even though the deviation of the magnetic moment of electrons from the Dirac value is relatively small, with the gyromagnetic ratio $g_\mathrm{electron}=-2-\alpha/\pi +\ldots$ we cannot assume that nonperturbative evaluation of electron-positron pair production allowing for this small correction is not required. We now clarify EM field regimes for which the non-perturbative treatment of electron magnetic moment for pair production matters, which turns out to be the magnetically dominated environment.

Using the reset~\req{gElectronreset} we recognize as noted below~\req{Rlim} the relevant parameter 
\begin{align} 
&\frac{g_\mathrm{electron}}{2}=-1-\frac{\alpha}{2\pi}\to \frac{g_1}{2}=1-\frac{\alpha}{2\pi}\;,\\
\label{chiB}
&\chi_b\equiv (1-|g_k|/2)\pi b/a=\alpha b/2a\;. 
\;
\end{align}
We now restate $R$,~\req{Rlim}, using~\req{chiB} and $\cos ( \pi l-x)= (-1)^l \cos x$ 
\begin{align} 
\label{Ralpha}
\!\!\!\!\!\!
&R_\mathrm{electron}
\\ \nonumber
&=
\frac{
\sum_{l=1}^\infty \frac{e^{-l\pi m^2/ea}} {l}\cos(l\alpha/2) 
\Big(e^{-l \chi_b} + e^{-l\pi b/a}\frac{\cosh(l \chi_b)}{\sinh(l\pi b/a)}\Big) 
}
{\sum_{l=1}^\infty \frac{e^{-l\pi m^2/ea}} {l} \coth(l\pi b/a)}
 \\ \nonumber
&=
\frac{\sum_{l=1}^\infty \frac{e^{-l\pi m^2/ea}} {l} \left[\coth(l\pi b/a)
 +\sum_{n=1}^\infty\alpha^nA_n(b/a;l)\right]}
{\sum_{l=1}^\infty \frac{e^{-l\pi m^2/ea}} {l} \coth(l\pi b/a) }
\;.
\end{align}
This is the final exact nonperturbative in ${\cal O}(\alpha)$ analytical result describing instability of the QED vacuum with regard to $e^-e^+$-pair production, using as reference the EHS result with $g=\pm 2$. The perturbative expansion in~$\alpha$ of the nominator term in~\req{Ralpha} creates the coefficient function $A_n$ shown to sixth order below 
\begin{align} 
\label{Rseries}
A_1 &= -\frac{l b/a}{2}\;,
 \\ \nonumber
A_2&=\frac{l^2}{ 8} \Big((b/a)^2-1\Big)\coth(l\pi b/a)\;,
 \\ \nonumber
A_3&=\frac{l^3}{48}\Big(3(b/a) -(b/a)^3\Big)\;,
 \\ \nonumber
A_4&=\frac{l^2}{384}\Big(1-6(b/a)^2+(b/a)^4\Big)\coth(l\pi b/a) \;,
 \\ \nonumber
A_5&= \frac{l^5}{3840}\Big(-5(b/a)+10(b/a)^3-(b/a)^5\Big)\;,
 \\ \nonumber
A_6&= \frac{l^6}{46080}\Big(-1+15l^4( b/a)^2-15(b/a)^4+(b/a)^6\Big) 
 \\ \nonumber
&\;\;\;\;\times\coth(l\pi b/a)
\;.
\end{align}
The series in powers of~$\alpha$ represents the individual contributions of the diagrams in~\rf{fig:sum}. 

The reference decay rate is seen in the denominator in~\req{Ralpha} and requires inclusion of the canceling common factor $\alpha ab/2\pi$, compare~\req{ImvactovacKrug}. Note that when the normalized electric field $\tilde a$ is sufficiently small only $l=1$ terms contribute in the~$\alpha$ series of the nominator. In \rf{fig:breakdown} we present results for $\tilde a=1$, and $\tilde a=1/10$ representative of weak but still functional $e^-e^+$-pair production. 
 
We show in \rf{fig:breakdown} the infinite order vertex summation alongside its perturbative expansion. Top frame is for $\tilde a=1$, while bottom for $\tilde a=1/10$. The solid line labeled \lq $\infty$\rq\ denotes the vertex correction summation to infinite order as seen in~\req{Ralpha}. The plots \lq $1,2,3,\ldots$\rq\ denote the perturbative summations in~\req{Rseries}, coefficients were shown to 6th order in $\alpha$. We see that for sufficiently large $b/a$ where $\chi_b>1$, the even power perturbative expansion in $\alpha$ breaks down while the odd-power becomes inaccurate. The nonperturbatively summed solution is needed to describe the physical $\mathfrak{Im}V^\mathrm{EHS}_g$ behavior when $\chi_b>1$. In sufficiently strong $b$-fields the perturbative summation is reliable only for $\chi_b<1$.

\begin{figure}
\centering
\includegraphics[width=0.99\columnwidth]{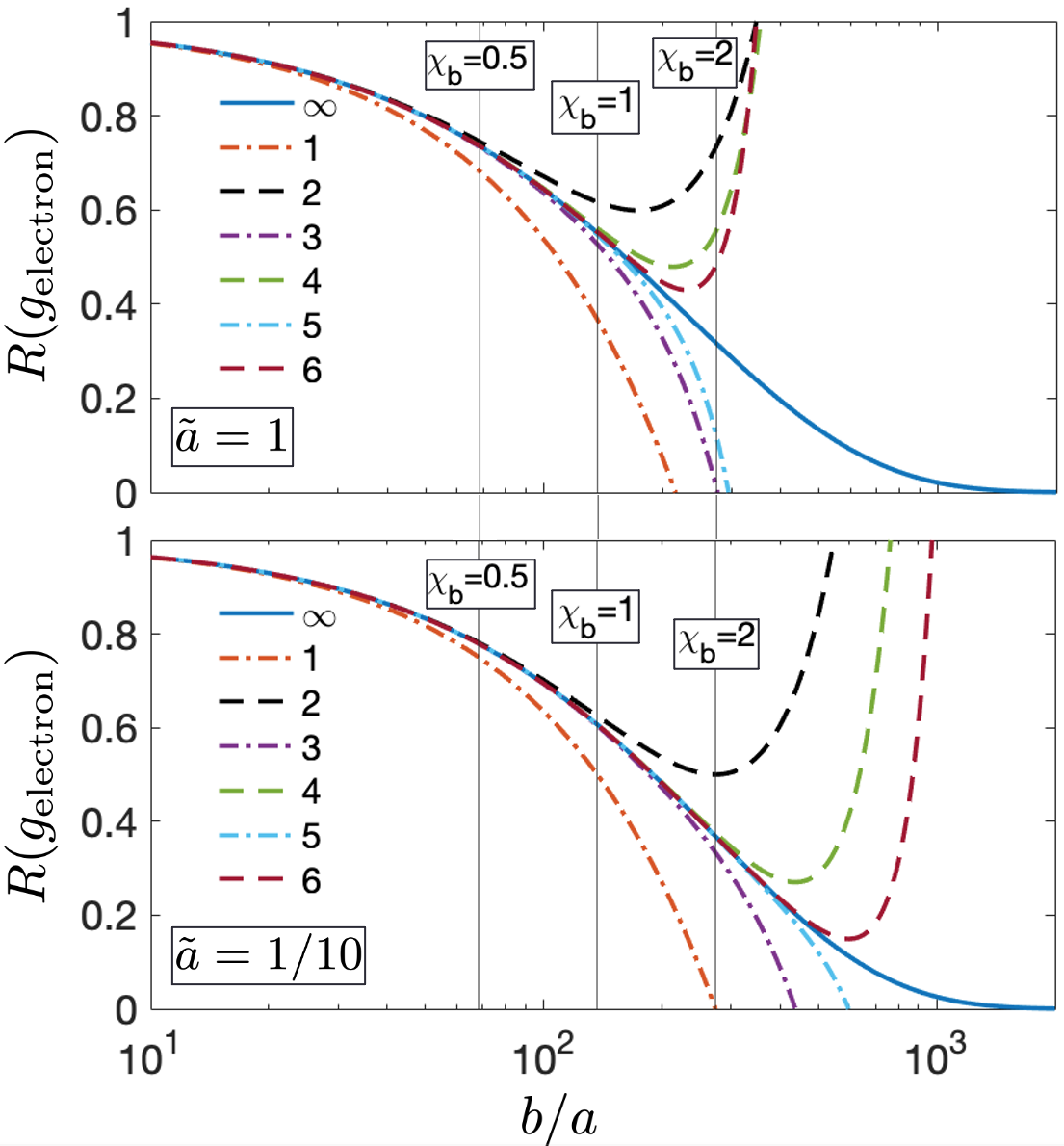} 
\caption{
\label{fig:breakdown} The suppression of the imaginary part of action $R$~\req{Ralpha} for $g_\mathrm{electron}=-(2+\alpha/\pi)$ (solid line) shown as a function of the (generalized) magnetic to electric field strength ratio $b/a$; top frame for $\tilde a=1$, bottom frame for $\tilde a=1/10$. Other lines: Perturbative expansion in powers of $\alpha$ of the analytic nonperturbative exact result. The orders in $\alpha$ are labeled. Vertical lines indicate the parameter $\chi_b$~\req{chiB} characterizing the magnetic dominance. }
\end{figure}

The results we presented are of interest in study of pair production in a magnetar environment. The magnetic fields in range $5<\tilde b<100$ are accompanied by electric fields which are at most subcritical~\cite{Kim:2022fkt}, offering a suitable environment for probing the nonperturbative $\chi_b$ regime seen in~\rf{fig:breakdown}. The search for particle production on magnetars is ongoing and remains an open question potentially testing QED~\cite{Ruffini:2009hg,Korwar:2017dio,Kim:2021kif}, considering both charged and neutral particles~\cite{Ferrer:2019xlr, Adorno:2021xvj}. Up to now studies of the electron vacuum response in magnetars have been based on evaluation of effective action at $g=\pm2$ and in following we extend this to physical values of $g$. 

However, in extreme environments of magnetars we cannot be sure that the magnetic moment of an electron is what it is in vacuum. Therefore we will consider the magnetic stabilizing effect varying the value of $g$ near to $g_\mathrm{electron}$. We note that the exponential suppression of particle production characterized by expansion parameter $\chi_b$ in~\req{chiB} depends only on the ratio $b/a$, and not the individual $a$ and $b$ strengths. The pair production suppression thus applies to a broad range of field configurations and values of $g$.

We demonstrate how the nonperturbative in $g$ suppression impacts this result by evaluating $R$ for different $\chi_b$ values relevant to magnetar fields. In~\rf{fig:supress} we plot $R$ from~\req{Rlim} for a small domain of $g$ centered on $g\sim2+\alpha/\pi$. We consider a subcritical electric-like $\tilde a=0.1$, and three magnetic $\tilde b$ values within the expected magnetar regime of 1-100 times the EHS critical field (\req{EBcritical})~\cite{Olausen:2013bpa, Enoto:2019vcg}. At $\pm g=2+\alpha/\pi$, we find for the case $\tilde a=0.1, \tilde b=5$ (blue, dashed, $\chi_b\sim1/5$) that the suppression effect is relatively small ($20\%$). The $g=\pm 2$ pair production is reduced by factor 2 for $\tilde a=0.1, \tilde b=25$ (red, dot-dashed, $\chi_b\sim1$). In the case $\tilde a=0.1, \tilde b=100$ (black, solid, $\chi_b\sim4$), the suppression is nearly two orders of magnitude. 

\begin{figure}
\centering
\includegraphics[width=0.77\columnwidth]{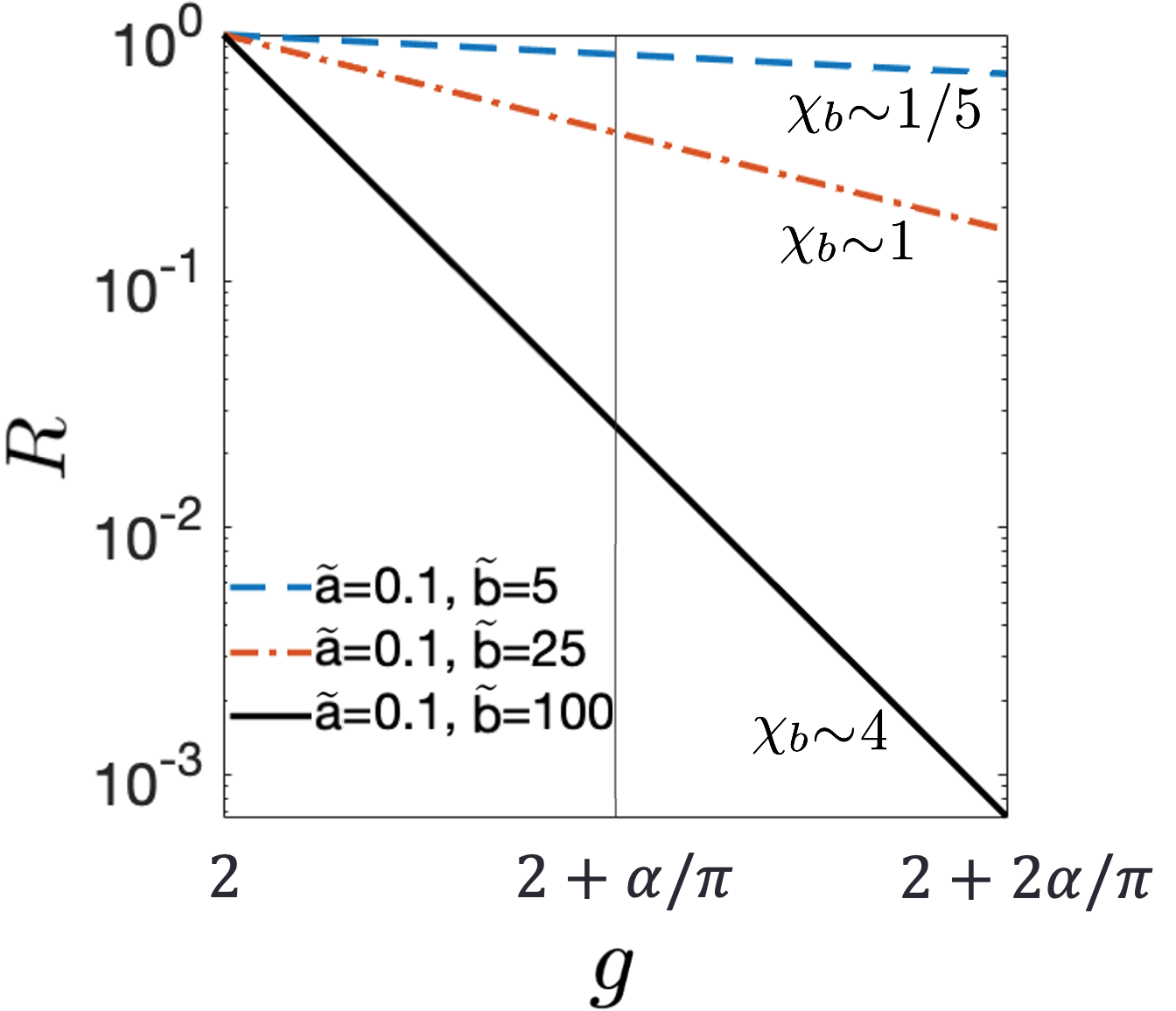}
\caption{$R$ from~\req{Rlim} plotted as a function of $g$, near to $|g_\mathrm{electron|}\sim 2+\alpha/\pi$ and the corresponding $\chi_b$ values from~\req{chiB} labeled. We present three different configurations of $\tilde a$ and $\tilde b$,~\req{abdef2}. 
\label{fig:supress} }
\end{figure}

To recognize this large suppression ab-initio consideration of $g\neq2$ is required. Had we considered $g=\pm 2$ exclusively, there would be a monotonically increasing (linear) in $b$ enhancing effect on particle production in magnetic dominated fields, quite opposite of the results we demonstrated. 

\subsection{Connection to non-abelian theory}
\label{Savvidy}

We comment on the behavior of the beta-function at $g$ values far from the singular points. Within the domain $|g|\leq 2$, the beta-function changes sign according to the charge-renormalization subtraction~\req{renormpart}, see~\cite{Angeles-Martinez:2011wpn, VaqueraAraujo:2012qa}. This result was extended to periodic domains for $|g|>2$~\cite{Rafelski:2012ui, Evans:2022fsu}:
\begin{align}
\label{Rasymptotically0}
\beta(e)=\mu\frac{\partial e}{\partial \mu}
=\frac{e^3}{12\pi^2}\Big(\frac{3g_k^2}8-\frac12\Big)
\;.
\end{align}
Interestingly,~\req{Rasymptotically0} is negative between the points where lines cross in~\rf{fig:KnE}, thus in domains of $g$ as follows
\begin{align}
\label{Rasymptotically}
\beta(e)<0\;,\qquad 
-\sqrt{4/3}+4k<g<\sqrt{4/3}+4k
\;.
\end{align}
Asymptotic freedom thus arises in such domains of the Abelian $V^\mathrm{EHS}_g$ formulation, allowing comparison with the non-Abelian Yang Mills vacuum~\cite{Savvidy:1977as}. 

We note that in these $g$ domains $\mathfrak{Im}V^\mathrm{EHS}_g$ is negative as seen in~\rf{fig:KnE}. Since both signs, $\beta$-function and pair instability, change sign in the asymptotically free domains~\req{Rasymptotically}, a study of QED vacuum structure may resolve the paradox of growing vacuum persistence probability that a negative imaginary part signals. Even so, recall that in the limit $b/a\to \infty$, the vacuum becomes completely stable even though pseudoscalar $\vec\mE\cdot\vec\mB$ is nonzero. This feature agrees with the recent finding that chromo-magnetic dominated fields with a nonvanishing pseudoscalar can be stable in the Savvidy vacuum state~\cite{Savvidy:2022}.

\subsection{Temperature representation}
\label{temprep}

The format of periodic in $g$ function $V^\mathrm{EHS}_g$ now available can be applied to extend prior work on the temperature representation of EHS action~\cite{Muller:1977mm, PauchyHwang:2009rz, Labun:2012jf}. First we consider the electric-dominated action. Like the proper time integration method for inserting $g\neq2$ in~\req{3.22b}, the temperature representation is on first sight restricted to the domain $|g|\leq 2$, see~\req{VeffE4} in~\rs{KGPfirst}, or Eq.\,(13) of~\cite{Labun:2012jf}. 

To extend this result to arbitrary $g$, we recall the conversion between effective action with $|g|>2$ to an equivalent form periodically reset to $|g_k|\leq 2$,~\req{gkreset}. This allows for a convergent result in both the proper time and the temperature representation integrands. Thus the prior result for $|g|\leq 2$ in Eq.\,(13) of Ref.~\cite{Labun:2012jf} requires only replacement $g\to g_k$ in order to describe all possible magnetic moments. Consequently, $V^\mathrm{EHS}_g$ for $g=2+ 4k$ values ($k=0,\pm1,\pm2,\ldots$) corresponds to the (spin inverted) bosonic distributions, while for $g=0+4k$ the representation is Fermionic.

We consider separately the case of magnetic-dominated fields. The procedure for deriving the corresponding temperature representation follows closely to the electric case summarized in~\rs{KGPfirst}. We start with the proper time integral form of $V^\mathrm{EHS}_g$ given by~\req{3.22c}, allowing for any $g$. In the vanishing electric field limit, with rotation of the integration contour $u\to-iu$,~\req{3.22c} becomes
\begin{align}
\label{3.22bB}
&V^\mathrm{EHS}_g=-\frac{1}{8\pi^2}\int_0^\infty\frac{du}{u^3}e^{-m^2u}
\Big(\frac{ ebu \cosh(\frac {g_k}2 ebu)}{\sinh(ebu)}-1\Big)
\;,
\end{align}
which we write in terms of the meromorphic expansion 
\begin{align} 
\label{meromorphicB}
&\frac{ ebs\cos(\frac {g_k}2 ebs)}{\sin(ebs)}-1=-\frac{e^2b^2u^2}3\Big(\frac{3g_k^2}8-\frac12\Big)
\\ \nonumber
&\;\qquad\qquad
- 2e^2b^2u^4\sum_{n=1}^\infty\frac{(-1)^n\cos(\frac{g_k}2 n\pi)}{n^2\pi^2(u^2+n^2\pi^2/e^2b^2)}
\;.
\end{align}
We plug~\req{meromorphicB} into~\req{3.22bB}, remove the charge renormalization contribution, and exchange summation with integration to obtain
\begin{equation} 
\label{VeffB}
V^\mathrm{EHS}_g
= 
\frac{e^2b^2}{4\pi^2}
\sum_{n=1}^\infty \int_{0}^{\infty}du\,u\,e^{-m^2u} 
\frac{(-1)^n\cos(\frac{g_k}2 n\pi)}{n^2\pi^2(u^2+n^2\pi^2/e^2b^2)}
\;.
\end{equation}
Substituting $u\to n\pi u/eb$ and exchanging summation with integration again,
\begin{align} 
\label{VeffB2}
V^\mathrm{EHS}_g=&\; 
\frac{e^2b^2}{8\pi^2}
 \int_{0}^{\infty}du\frac {2u}{u^2+1} 
 \\ \nonumber
&\;\times
\sum_{n=1}^\infty e^{-n\pi m^2u/eb} 
\frac{(-1)^n\cos(\frac{g_k}2 n\pi)}{n^2\pi^2}
\;.
\end{align}
Integrating~\req{VeffB2} by parts, 
\begin{equation} 
\label{VeffB3}
V^\mathrm{EHS}_g = 
v
\int_{0}^{\infty}\!\!\!du \ln[u^2+1]
\sum_{n=1}^\infty e^{-n\beta u} 
\frac{(-1)^n\cos(\frac{g_k}2 n\pi)}{n\pi}\;,
\end{equation}
where now
\begin{equation} 
\beta=\pi m^2/eb\;,\qquad v=\frac{m^4}{8\pi^2\beta}
\;.
\end{equation}
Summing~\req{VeffB3} over $n$ we obtain 
\begin{align} 
\label{VeffB4}
V^\mathrm{EHS}_g
= &\;
-v
\int_{0}^{\infty} du \ln[u^2+1]
\frac{1}{2}\sum_\pm \ln[1+e^{-\beta u}e^{\pm i\frac{g_k}2 \pi}]
\;.
\end{align}

Comparing the magnetic field action~\req{VeffB4} to the electric case Eq.\,(13) of~\cite{Labun:2012jf} (\req{VeffE4} in~\rs{KGPfirst} here), the $g$-dependent logarithmic terms on the RHS are identical within the domain $|g|\leq 2$. Thus in this $g$ domain the two expressions obey the same statistical model representation. Given that our conversion $g\to g_k$ is equal for electric and magnetic fields, the periodic $|g|>2$ extensions of these two statistical representations are also equal. The one difference is between the spectral function terms describing density of the virtual particles: $\ln[u^2+1]$ appears in the $\mB$-dominated integrand, compared to $\ln(u^2-1+i\varepsilon) $ in the $\mE$ field case.

\section{Summary, Conclusions and Outlook} 
\label{summary2}

We generalized the Euler-Heisenberg-Schwinger (EHS) effective action to arbitrary values of the gyromagnetic ratio $g$. We have demonstrated a cusp singularity in the vicinity of $g=\pm 2, \pm 6, \ldots $. For arbitrary quasi-constant field configurations we have shown periodicity in $V^\mathrm{EHS}_g$ as a function of $g$ recognized before for the pure electric $\mE$~\cite{Evans:2022fsu}, and pure magnetic $\mB$ field cases~\cite{Rafelski:2012ui}. 

The nonperturbative in $g$ singular behavior of $V^\mathrm{EHS}_g$ in the presence of an anomalous magnetic moment $g\ne 2$ was conjectured in our prior work~\cite{Rafelski:2012ui, Evans:2022fsu}. In this work using the Bogoliubov coefficient summation method~\cite{Nikishov:1979ez, Kim:2008yt} we were able prove in \rs{pseduoNonpert} this singular behavior dominated by nonvanishing pseudoscalar $\vec\mE\cdot\vec\mB$. This feature escaped prior attention since considered for pure electric $\mE$, or pure magnetic $\mB$ fields, the effective action $V^\mathrm{EHS}_g$ is smooth and differentiable at $g=\pm 2$. This exact summation procedure resolves the effective action for both $|g|\le2$ and $|g|>2$, where analytical continuation between the two domains is otherwise not possible due to the cusp.

We have shown that the sharpness of the cusp in $\mathfrak{Im}V^\mathrm{EHS}_g$ at $g=\pm 2$ is dependent on the EM fields in a nonperturbative fashion, and occurs for nonzero $\vec\mE\cdot\vec\mB$, based on the nonperturbative discontinuity in $dV^\mathrm{EHS}_g/dg$ shown in~\req{Imvactovac9},~\rs{CuspShape}. We believe that the importance of nonvanishing pseudoscalar $\vec\mE\cdot\vec\mB$ is echoed by the relatively strong coupling of two photons to the singlet pseudoscalar $0^-$-para-positronium and the related fast decay channel $\tau_s=0.124$\,ns; to be compared to $\tau_t=142$\,ns for the triplet ortho-positronium coupling to odd number of photons.

We have explored some of the nonperturbative properties of $V^\mathrm{EHS}_g$. Most interesting is the cusp singularity for magnetically dominated fields $\mathfrak{Im}V^\mathrm{EHS}_g$ capability to heavily suppress the pair production, see~\rf{fig:KnE} in~\rs{CuspShape}. We presented explicit dependence of this suppression effect exploring several important values of EM fields $\mE\to a$, $\mB\to b$,~\rf{fig:KnE} in~\rs{CuspShape}, confirming the conjectured results presented in Ref.\,\cite{Evans:2018kor}. Considering proportionality of the magnetic moment $\mu\propto g/m$ and viewing our results as a function of $\mu$ rather than $g$ we conjecture equivalence of our results to an effective mass modification,~\req{mshift}. In this case our result is reminiscent of the analysis of higher order loop contributions to the imaginary part of EHS action carried out in~\cite{Affleck:1981bma, Lebedev:1984mei}. However this was applied to electrically dominated fields, while our present work focuses on magnetically dominated environments.

We have identified a smallness parameter $\chi_b$,~\req{chiB} describing at which value of $b/a$ the nonperturbative $g$-modification of the EHS action is significant, see~\rf{fig:breakdown} in~\rs{Stability}. In the context of a systematic perturbative expansion we recall the Ritus-Narozhny conjecture~\cite{Ritus:1970, Narozhnyi:1979at}, where the parameter $\alpha\chi^{2/3}$ ($\chi=(e/m^3)\sqrt{(-F^{\mu\nu}p_\nu)^2}\to e\mE/m^2|_\mathrm{rest\ frame}$) is considered to govern the breakdown of perturbative QED, spurring exploration of convergence of higher order radiative QED corrections~\cite{Fedotov:2016afw, Mironov:2020gbi, Edwards:2020npu, Torgrimsson:2021zob, Dunne:2021acr, Heinzl:2021mji, Mironov:2022jbg, Sasorov:2022vqk, Fedotov:2022ely}. Our result demonstrates parallels with this study, where the the nonperturbative in $\vec\mE\cdot\vec\mB$ magnetically dominated EM fields present an entirely different environment in which we identified strong suppression of particle production.

We have presented explicit dependence of this suppression effect on the ratio $b/a$, allowing for direct application of our results to EM fields relevant to astrophysical environments such as magnetars, and heavy ion collisions in which the $\mB$ field dominates near-critical $\mE$ fields~\cite{Ruffini:2009hg,Korwar:2017dio,Kim:2021kif}. It is important to recognize the cusp in $g$ in current perturbative schemes, since the anomalous magnetic moment can suppress the particle production rate by orders of magnitude, see~\rf{fig:supress} in~\rs{Stability}. This nonperturbative $g$-dependence is a step towards addressing the question as to whether magnetar fields generate pair production or are pair-stable environments.

The singular behavior at $g=\pm 2$ for quasi-constant fields of any strength leads us to the question more generally: Could there be higher order modifications of the conventional perturbative QED expansion which is carried out at $g=\pm 2$ reflecting on this singular behavior in presence of external fields? This is probably so: The perturbative series for $g$ in QED relies on the evaluation of the energy change of a particle in presence of an external EM field and this is exactly what we have done using the external field EHS method for quasi-constant fields. 

Since the effective action dependence on $g$ is nonperturbative for certain external EM field configurations, a perturbative series defining $g$ should be formulated allowing for singular behavior; even a small deviation from the Dirac equation value $g=\pm 2$ can have a significant effect. Addressing this situation in the context of actual precision experimental environment is perhaps the most important open question arising from our work. Answer to this question requires entirely different technical methods and is completely outside the scope of this work.

We described the singular behavior of the imaginary part of the effective action $\mathfrak{Im}V^\mathrm{EHS}_g$ as a function of $g$ considering the pair production rate in~\req{Imvactovac9}. The singular properties of the full effective action $V^\mathrm{EHS}_g$ require another consideration beyond the scope of this work: $g$ appears associated with the magnetic field $b$ since $g$ acts as a spin - field coupling. Thus a singular behavior in $g$ seen in~\req{Imvactovac9} indicates also singular behavior of $V^\mathrm{EHS}_g$ as function of $b$. Presence of a cusp as a function of $gb$ would appear as a discontinuity in the magnetic susceptibility. We conclude that our results may be indicating presence of a 2nd order phase transition in QED in presence of magnetically dominated strong fields.

Our QED result for strong fields considered with variable magnetic moment can mimic asymptotic freedom of strong interactions and there are some parallels of our work with the those usually associated with vacuum structure in QCD. For example $\mathfrak{Im}V^\mathrm{EHS}_g$ is suppressed in certain domains of $g$ in which also asymptotic freedom arises in our Abelian theory,~\req{Rasymptotically}. This feature parallels recent results of Savvidy~\cite{Savvidy:2022} who demonstrated that the asymptotically free Yang Mills Lagrangian is stable in (chromo) magnetic dominated fields, allowing for the presence of nonvanishing (chromo) $\vec\mE\cdot\vec\mB$ field configurations. 

To further compare our QED result with features of QCD vacuum requires understanding what ratio $\mB/\mE$ is needed to stabilize the vacuum. This condition is clearly met in the here adopted strong field diagram resummation only in the $\mB/\mE\to\infty$ limit: When $\mE$ and $\mB$ are of the same order there remains an exponentially suppressed in $b/a$, see~\req{Rlim} nonzero imaginary part. Further study of this interesting result may require consideration of resummation of infinitely many higher order corrections to the effective action.

We have extended the temperature representation of the $V^\mathrm{EHS}_g$ effective action for all $|g|$, extending prior work based on $g=\pm 2$~\cite{Muller:1977mm, PauchyHwang:2009rz} and $|g|\leq 2$~\cite{Labun:2012jf}. We obtained a result for pure magnetic fields, which exhibits the same statistical representation as the electric case. Further exploration is needed to understand the role of the pseudoscalar $\vec\mE\cdot\vec\mB$ contribution. An indication that same statistical form arises for nonvanishing $\vec\mE\cdot\vec\mB$ can be found in~\rs{CuspShape}. The sharpness of the $\vec\mE\cdot\vec\mB$-dependent cusp~\req{Imvactovac9}, depends on the term $(e^{\pi m^2/ea}-1)^{-1}$, which takes on a Bose distribution at $g=\pm 2$ (inverted spin statistics) in agreement with~\cite{Muller:1977mm}.

Our nonperturbative results differ from the work of Ritus~\cite{Ritus:1975cf} in that we have summed exclusively the vertex diagrams in~\rf{fig:sum}, to nonperturbative infinite loop order. Inspecting~\rf{fig:sum} further, one notices that our approach does not fully account for all possible perturbative corrections, as it misses diagrams where an internal photon line crosses the Fermion loop isolating at least two external photon lines to the right and left. Such contributions arise in second and higher orders in the external EM field from self-energy corrections to the Fermion propagator~\cite{Ritus:1970, Jancovici:1970ep, Newton:1971pq, Constantinescu:1972qe, Tsai:1974id, Narozhnyi:1979at, Morozov:1981pw, Loskutov:1981bk, Gusynin:1998nh, Machet:2015swa}. They produce field-dependent corrections to mass and $g$~\cite{Ferrer:2015wca, DiPiazza:2021szp}, and in closed form give the leading Ritus two-loop and higher order~\cite{Ritus:1975cf, Dunne:2004xk, Huet:2017ydx, Huet:2018ksz, Dunne:2021acr} corrections to the EHS action.   

This leads to the question: will the cusp in $g$ arising from the infinitely summed vertex diagrams persist in a truly complete QED solution? The following evidence strongly suggests the cusp will persist: The sharpness of the cusp is governed by the ratio in EM invariants $b/a$ as it appears in~\req{chiB}-\req{Rseries} where we define the expansion in $\alpha b/a$ parameter,  independent of the individual $a$, $b$ values. In contrast, the higher order self-energy corrections take on a different EM field dependence, where individual values of $b$ and $a$ matter too~\cite{Ritus:1975cf, Ritus:1970}. Given the structure of the mathematical expressions we do not expect that an \lq\lq opposite sign cusp\rq\rq\ could arise to cancel the vertex diagram effect at all field strengths. Dependence of the QED singularity on the pseudoscalar $P=\vec\mE\cdot\vec\mB$ is a very intriguing feature.

We further note that while the $V^\mathrm{EHS}_g$ result considers exclusively the (irreducible) vertex contributions to the EHS action, there are also reducible diagrams recently found to be nonvanishing in constant EM fields: As effective action~\cite{Gies:2016yaa, Karbstein:2017gsb, Karbstein:2019wmj} and propagator~\cite{Ahmadiniaz:2017rrk, Edwards:2017bte} contributions. In these works it was shown that the reducible corrections dominate the irreducible contributions, based on the strong field asymptotic behavior of $S=(\mE^2-\mB^2)/2$-dominated effective action. The dependence on $P=\vec\mE\cdot\vec\mB$ has not been yet been considered, and can influence the strong field asymptotic.

To conclude: We have obtained the generalized EHS effective action accounting for anomalous $g\neq2$ and included the effect of pseudoscalar $\vec\mE\cdot\vec\mB$. Nonperturbative phenomena are uncovered in resummed expressions: Radiative corrections, previously assumed to be small within perturbative QED context, are the dominant contributions for certain EM field configurations. Our result could be a step toward novel understanding of the singular properties in QED noted for example by Dyson~\cite{Dyson:1952tj} and K\"all\'en~\cite{Kallen:1957ib, Kallen:1972pu}. Our results also provide means to identify parallels of strong field vacuum QED phenomena with the strongly interacting QCD vacuum.


\end{document}